\documentclass[preprint,aps,prd,superscriptaddress,preprintnumbers,letterpaper,natbib,floatfix,nofootinbib]{revtex4-1}

\usepackage[normalem]{ulem}
\usepackage{graphicx} 
\usepackage{axodraw4j}
\usepackage{amsmath}
\usepackage{amssymb}
\usepackage{caption}
\usepackage{subcaption}

\usepackage{ragged2e}
\DeclareCaptionJustification{justified}{\justifying}
\usepackage{tikz,xcolor,hyperref}

\definecolor{lime}{HTML}{A6CE39}
\DeclareRobustCommand{\orcidicon}{\hspace{-1mm}
	\begin{tikzpicture}
		\draw[lime, fill=lime] (0,0) 
		circle [radius=0.16] 
		node[white] {{\fontfamily{qag}\selectfont \tiny \,ID}};
		\draw[white, fill=white] (-0.0525,0.095) 
		circle [radius=0.007];
	\end{tikzpicture}
	\hspace{-3mm}
}

\foreach \x in {A, ..., Z}{\expandafter\xdef\csname orcid\x\endcsname{\noexpand\href{https://orcid.org/\csname orcidauthor\x\endcsname}
		{\noexpand\orcidicon}}
}

\begin{document}

\title{\large Singlet-doublet dark matter revisited}
\author{Prudhvi N.~Bhattiprolu\orcidA}
\email{prudhvi.bhattiprolu@fi.infn.it}
\affiliation{INFN, Sezione di Firenze, Via Giovanni Sansone 1, 50019 Sesto Fiorentino, Italy}
\affiliation{Leinweber Center for Theoretical Physics, Department of Physics,\\ University of Michigan, Ann Arbor, MI 48109, USA}
\author{Evan Petrosky\orcidB}
\email{epetros@umich.edu}
\affiliation{Leinweber Center for Theoretical Physics, Department of Physics,\\ University of Michigan, Ann Arbor, Michigan 48109, USA}
\author{Aaron Pierce\orcidC}
\email{atpierce@umich.edu}
\affiliation{Leinweber Center for Theoretical Physics, Department of Physics,\\ University of Michigan, Ann Arbor, MI 48109, USA}

\begin{abstract}
The singlet-doublet model is an economical model of weakly interacting dark matter.  We revisit it in light of improved dark matter direct detection limits.  We characterize the now well-defined regions of remaining parameter space with suppressed direct detection cross sections and discuss features of the spectrum accessible at the Large Hadron Collider.  We discuss when and how parameters in these special regions might be realized as the result of renormalization group evolution when starting with generic ultraviolet initial conditions.
\end{abstract}

\maketitle
\preprint{LCTP-25-06}
\tableofcontents
\newpage

\section{Introduction}\label{sec:intro}
With the improvement in limits on the dark matter (DM) direct detection cross section from the LUX-ZEPLIN (LZ) experiment in 2024 \cite{LZCollaboration:2024lux}, it is timely to consider the status of the weakly interacting massive particle (WIMP) paradigm.

A particularly simple model of the WIMP is the so-called singlet-doublet dark matter model  \cite{Arkani-Hamed:2005zuc, Kearney:2016rng, Cohen:2011ec, Cheung:2013dua, DEramo:2007anh, Mahbubani:2005pt, Enberg:2007rp, Abe:2014gua, Calibbi:2015nha, Freitas:2015hsa,Banerjee:2016hsk, Egana-Ugrinovic:2017jib, LopezHonorez:2017zrd, Esch:2018ccs, Arcadi:2018pfo}.  For reviews that discuss the singlet-doublet model among a range of other WIMP possibilities, see \cite{Arcadi:2024ukq, Arcadi:2019lka, Arcadi:2017kky, Cirelli:2024ssz}, see also \cite{Paul:2024prs}. This model includes a $SU(2)_L$ vectorlike fermion doublet that mixes with a Majorana fermion gauge singlet; that is, it has particle content in common with the Bino-Higgsino sector of the minimal supersymmetric Standard Model (MSSM). However, the Yukawa couplings present in the model are not constrained by supersymmetry, so they need not be proportional to gauge couplings.  As such, this model is a generalization of the neutralino DM of the MSSM.  Some regions of the parameter space reproduce the phenomenology of dark matter in split supersymmetry \cite{Arkani-Hamed:2004ymt, Giudice:2004tc,Pierce:2004mk, Cheung:2012qy, Ibe:2023dcu, Nagata:2014wma, Martin:2024pxx, Martin:2024ytt}. Moreover, while the renormalization group evolution of the gauge couplings is not identical to the MSSM, the introduction of the $SU(2)_L$ doublet fermions substantially improves the quality of gauge coupling unification with respect to the Standard Model (SM) \cite{Arkani-Hamed:2005zuc}.   The purpose of studying this model is twofold.  First, it is a simple model for DM, and as such is of interest in its own right.  Second, because this model generates a thermal DM abundance via interactions with the $W$, $Z$, and Higgs bosons, studying it can give insight into the status of the WIMP paradigm more generally. 

In this paper, we discuss the status of the singlet-doublet model in light of recent direct detection limits.  We identify parameter space as yet unconstrained by these bounds and also the parameter space that will remain challenging to probe with future direct detection experiments. We will show how current direct detection bounds allow sharp characterization of the remaining parameter space. The model consists of three neutral Majorana fermions (the lightest of which is the DM) and one charged fermion. Because the parameter space is well constrained, it is possible to make predictions for the mass spectrum of these fermions.  This is in particular true for the parameter space that has DM within kinematic reach of the Large Hadron Collider (LHC).   This has implications for potential LHC signals.

The structure of this paper is as follows.  In Sec.~\ref{sec:background}, we review the details of the singlet-doublet model.  We highlight the most relevant couplings for direct detection and for the determination of the relic density.  Coannhilation \cite{Griest:1990kh} plays an important role. In Sec.~\ref{sec:results} we provide numerical results where we scan over the parameter space of this model, with emphases on i) identifying regions that will remain stubbornly difficult to probe via direct detection, and ii) the interplay between direct detection and the DM relic abundance.  We discuss the relevance of LHC measurements as a complementary probe.  In Sec.~\ref{sec:focusing} we explore whether renormalization group (RG) evolution might explain why the masses and couplings  would reside in a region with low direct detection cross sections.  Finally, we conclude.

\section{Singlet-doublet Dark Matter}\label{sec:background}

The singlet-doublet fermion model introduces three left-handed Weyl fermions which we call $S$, $D$, and $\overline{D}$ (the bar is part of the name and does not denote any sort of conjugation).  These three fermions are in the
$({\bf 1}, {\bf 1}, 0)$,
$({\bf 1}, {\bf 2}, -\frac{1}{2})$, and
$({\bf 1}, {\bf 2}, \frac{1}{2})$
representations of
the SM gauge group
$SU(3)_c \times SU(2)_L \times U(1)_Y$ respectively.  The states are odd under a $\mathbb{Z}_2$ symmetry while the SM states are all even under the same $\mathbb{Z}_2$.  This discrete symmetry ensures the stability of the lightest new particle (i.e. our dark matter candidate). 

We denote the $SU(2)_L$ components of the doublets $D$ and $\overline{D}$ as 
\begin{equation}
    D = \begin{pmatrix}
        N \\
        C
    \end{pmatrix}
,\qquad
    \overline{D} = \begin{pmatrix}
        \overline{C} \\
        \overline{N}
    \end{pmatrix}
,
\end{equation}
where $N$ and $\overline{N}$ are neutral, and $C$ and $\overline{C}$ are charged under $U(1)_\text{EM}$.
As mentioned earlier, the field content of the model is identical to that of the Bino-Higgsino sector of the MSSM.  The doublets correspond to the two Higgsinos, while the singlet maps to the Bino.

The kinetic and gauge terms in the Lagrangian are
\begin{equation}
\begin{aligned}   
    \mathcal{L}_{gauge} = i S^{\dagger}\overline{\sigma}^{\mu} \partial_{\mu} S + i D^{\dagger}\overline{\sigma}^{\mu}\nabla_{\mu} D + i \overline{D}^{\dagger}\overline{\sigma}^{\mu}\nabla_{\mu}\overline{D}
    ,
\end{aligned}
\end{equation}
where $\nabla_{\mu}$ is the covariant derivative.  After electroweak symmetry breaking, these terms result in couplings between the SM $Z$, $W$, and photon and the neutral and charged $SU(2)_L$ components of the doublets.

The mass and Yukawa couplings of the new fields are
\begin{equation}
\begin{aligned}   
    \mathcal{L}_{mix} = - \frac{1}{2} M_S S^2 - M_D D \overline{D}
    - y_1 D H S - y_2 H^{\dagger} \overline{D} S + H.c.,
\end{aligned}
\end{equation}
where $H \sim ({\bf 1}, {\bf 2}, \frac{1}{2})$ is the SM Higgs doublet, whose neutral component contains the physical Higgs boson $h$,
and $M_S, M_D, y_1, y_2$ are the four new parameters introduced in this singlet-doublet model.  Via field redefinitions,  without loss of generality, we may consider the case where $M_S, M_D, y_1$ are positive and real, and the remaining physical complex $CP$ violating phase is solely contained in $y_2$.  For this analysis we specialize to the case where this $CP$ violating phase vanishes and so $y_2 \in \mathbb{R}$. For extensions to the case of $CP$ violation see, e.g., \cite{Abe:2014gua,Abe:2019wku, Egana-Ugrinovic:2017jib}.  The case with $CP$ violation can have implications for electric dipole moment searches. For $y_1^2 + y_2^2 \rightarrow \frac{g^{\prime 2}}{2}$, we  reproduce the Higgs decoupling limit of the MSSM with a decoupled Wino  by the following identifications \cite{Calibbi:2015nha}
\begin{equation}
    M_D \rightarrow -\mu, \qquad M_S \rightarrow M_1, \qquad -\frac{y_2}{y_1} \rightarrow \frac{1}{\tan \beta}. 
\end{equation}
After the SM Higgs field takes on its vacuum expectation value $v$, the neutral components of the doublets and the singlet mix.

In terms of the gauge eigenstate basis of the neutral states $\psi^0 \equiv (S, N, \overline{N})$, the mass mixing is given by
\begin{equation} \label{eq:Lagrangian}
    \mathcal{L}_{mass} = -\frac{1}{2}(\psi^0)^T \mathcal{M}_N \psi^0 - C (-M_D)\overline{C}
    ,
\end{equation}
where
\begin{equation} \label{eq:mixingmatrix}
     \mathcal{M}_N = 
    \begin{pmatrix}
       M_S & \frac{y_1 v}{\sqrt{2}} & \frac{y_2 v}{\sqrt{2}} \\
       \frac{y_1 v}{\sqrt{2}} & 0 & M_D \\
       \frac{y_2 v}{\sqrt{2}} & M_D & 0
    \end{pmatrix}
    ,
\end{equation}
and $v \simeq 246$ GeV is the SM Higgs vacuum expectation value.

Diagonalization of this mass matrix yields three neutral Majorana mass eigenstates which (borrowing notation from the MSSM) we denote as $\chi^0_i$ with masses $m_i$  ($i =1,2,3$).  We order the neutral mass eigenstates from smallest to the largest. The lightest neutral particle $\chi^0_1$ is the DM candidate.  Denoting the diagonalization matrix by $V$, we have

\begin{equation}
    \text{diag}(m_1, m_2, m_3) = V^{*}\mathcal{M}_NV^{\dagger},
\end{equation}
and 
\begin{equation}
    \chi^0_i = V_{ij}\psi^0_j,
\end{equation}
where repeated indices are summed.  The elements of the diagonalization matrix $V_{ij}$ appear in the couplings of the neutral states with the SM Higgs and gauge bosons, see below. For the charged states, there is a single Dirac fermion with mass $m_{\pm}$ (at tree level, this is just equal to $M_D$).

\subsection{Couplings and blind spots}\label{sec:bs}

The dark sector particles couple to the SM exclusively via the SM gauge and Higgs bosons.
In terms of four-component neutral Majorana fermions
$\Psi_{\chi^0_i} \equiv (\chi^0_i \>\> \chi^{0 \> \dagger}_i)^T$
and the charged Dirac fermion
$\Psi_{\chi^-} \equiv (C \>\> -\overline{C}^\dagger)^T$, 
referred to as $\chi^0_i$ and $\chi^-$ in the following,
couplings to the photon $A_\mu$, $Z$ boson $Z_\mu$,  $W$ bosons $W^\pm_\mu$, and the physical Higgs boson $h$
are:
\begin{equation}
\begin{aligned}   
    \mathcal{L} & \supset
    e A_\mu \overline{\chi^-} \gamma^\mu \chi^- - \frac{e}{s_W c_W} \left(s_W^2 - \frac{1}{2}\right) Z_\mu \overline{\chi^-} \gamma^\mu \chi^-\\
    & - g^V_{W^+ \chi^- \chi^0_i} W^+_\mu \overline{\chi}^0_i \gamma^\mu \chi^-
    - g^A_{W^+ \chi^- \chi^0_i} W^+_\mu \overline{\chi}^0_i \gamma^\mu \gamma_5 \chi^- + c.c.\\
    & - \frac{1}{2} \text{Re}(g_{Z \chi^0_i \chi^0_j}) Z_\mu \overline{\chi}^0_i \gamma^\mu \gamma_5 \chi^0_j
    + \frac{i}{2} \text{Im}(g_{Z \chi^0_i \chi^0_j}) Z_\mu \overline{\chi}^0_i \gamma^\mu \chi^0_j\\
    & - \frac{1}{2} \text{Re}(g_{h \chi^0_i \chi^0_j}) h \overline{\chi}^0_i \chi^0_j
    + \frac{i}{2} \text{Im}(g_{h \chi^0_i \chi^0_j}) h \overline{\chi}^0_i \gamma_5 \chi^0_j,
    \label{eq:SMcouplings}
\end{aligned}
\end{equation}
where
$e$ is the electromagnetic coupling, $s_W (c_W)$ is the (co)sine of the weak mixing angle, and
\begin{equation}
\begin{aligned}
g^V_{W^+ \chi^- \chi^0_i} & = \frac{e}{2\sqrt{2} s_W} \left(V_{i 3}^* +  V_{i 2}\right),\\
g^A_{W^+ \chi^- \chi^0_i} & = \frac{e}{2\sqrt{2} s_W} \left(V_{i 3}^* -  V_{i 2}\right),\\
g_{Z \chi^0_i \chi^0_j} & = \frac{e}{2 s_W c_W} \left(V_{i 3} V_{j 3}^* - V_{i 2} V_{j 2}^*\right),\\
g_{h \chi^0_i \chi^0_j} & = \sqrt{2} \left(y_1 V_{i 2}^* V_{j 1}^* + y_2 V_{i 3}^* V_{j 1}^*\right).
\label{eq:SMcouplings2}
\end{aligned}
\end{equation}
Here, $g_{Z \chi^0_j \chi^0_i} = g^*_{Z \chi^0_i \chi^0_j}$, and $g_{h \chi^0_i \chi^0_j}$ is unsymmetrized in $i$ and $j$ (i.e., $g_{h \chi^0_j \chi^0_i} \ne g_{h \chi^0_i \chi^0_j}$ for $i \ne j$).

We provide details on the couplings and explicitly evaluate them in terms of the underlying model parameters in relevant limits in Appendix \ref{sec:appendix}.  The coupling of the DM candidate $\chi^0_1$  to the Higgs boson generates the effective operator $(\overline{\chi}_1^0 \chi_1^0)(\overline{q}{q})$,
and thus controls the spin-independent (SI) direct detection cross section. The DM coupling to the $Z$ boson is responsible for the operator ($\overline{\chi}_1^0 \gamma_{\mu} \gamma_{5} \chi_1^0)( \overline{q} \gamma^{\mu} \gamma_5{q})$, and thus spin-dependent (SD) direct detection.  

For certain areas of parameter space,
the couplings $g_{Z \chi^0_1\chi^0_1}$ and $g_{h \chi^0_1\chi^0_1}$
vanish.  These points are called blind spots \cite{Cohen:2011ec}; see \cite{Cheung:2012qy} for a lucid discussion for the analogous phenomenon in the context of the MSSM; see also \cite{Badziak:2015exr,Badziak:2017uto} for an analogous phenomenon in the singlino-Higgsino sector of the NMSSM. When this occurs,  the corresponding direct detection cross sections also vanish at tree level. The blind spots occur when the numerators of Eq.~(\ref{eq:Zcoupling}) and Eq.~(\ref{eq:Higgscoup}) vanish.  Requiring  the model inhabit blind spot(s) imposes conditions on the underlying parameters of the model.  In this work, we discuss these conditions as a restriction on the ratio of the Yukawa couplings.  So, when we refer to a ``blind spot" we refer to a specific coupling ratio for fixed values of the mass parameters.   

The DM coupling to the $Z$ boson $g_{Z\chi^0_1\chi^0_1}$ vanishes when
\begin{equation} \label{eq:ZBS}
    \frac{y_2}{y_1} = \pm 1
    ,
    \qquad \text{($Z$ blind spots)}
\end{equation}
independent of the mass parameters $M_S$ and $M_D$.
This condition also restores a custodial $SU(2)$ symmetry in the dark sector.

For $0 < M_S < M_D$, the coupling to the Higgs boson $g_{h\chi^0_1\chi^0_1}$ vanishes when $2 y_1 y_2 M_D + (y_1^2 + y_2^2) M_S = 0$ \cite{Cohen:2011ec} --  this condition can be derived by Higgs boson low energy theorems -- it says the DM mass does not depend on the Higgs vacuum expectation value.   It implies
\begin{equation} \label{eq:HBS}
     \frac{y_2}{y_1} = -\frac{M_S}{M_D}\left(1\pm\sqrt{1-\left(\frac{M_S}{M_D}\right)^2}\right)^{-1}
     .
     \qquad \text{($h$ blind spot; $M_S < M_D$)}
\end{equation}
One of the solutions is redundant; it may be obtained through a relabeling of the model parameters, i.e., by the exchange $y_1 \leftrightarrow y_2$.
Here, without loss of generality, we take the solution with the plus sign in front of the square root, such that $0 < -y_2/y_1 < 1$ for $0 < M_S/M_D < 1$.

When $0<M_D<M_S$,
the couplings of the dark matter to the Higgs boson ($g_{h\chi^0_1\chi^0_1}$) and $Z$ ($g_{Z\chi^0_1\chi^0_1}$) can be
simultaneously eliminated by the condition
\begin{equation} \label{eq:DoubleBS}
    \frac{y_2}{y_1} = -1
    .
    \qquad \text{(double blind spot; $M_D < M_S$)}
\end{equation}
When $M_{D} < M_{S}$, the DM is predominantly doublet in nature, and it behaves much like the Higgsino dark matter in the MSSM.

These blind spots play a role in understanding the remaining allowed parameter space for the model, as proximity to the blind spot(s) is one way to avoid the increasingly stringent direct detection bounds. The blind spots derived above are tree-level relations.  It is natural to ask whether  inclusion of loop corrections spoils any chance of suppressing the cross section beyond the reach of experiments. This is not the case.   If the tree-level cross section vanishes due to an accidental relation between the model parameters and is not due to a symmetry, Ref. \cite{Han:2018gej} (in the related MSSM context) demonstrated that a blind spot will typically exist despite the loop correction, just in a location slightly shifted from the tree-level blind spot. For the case where the tree-level cross section vanishes identically due to a symmetry, then the expected size of the direct detection cross section would be given by the loop-induced one.  However, for our parameter space of interest, this loop-induced cross section is expected to be below the neutrino fog \cite{Chen:2018uqz, Chen:2019gtm} and thus does not impact the results of this analysis. In particular, it has no bearing on identification of points that will remain stubbornly difficult to probe via direction detection well into the future.


After imposing that the couplings relevant for direct detection (nearly) vanish, one might wonder whether it is still possible to achieve the observed dark matter relic density. Conceivably, turning off these  dark matter couplings could force the dark matter to interact too weakly with the Standard Model, and thus freeze-out to a too large abundance. But even when the couplings of the $Z$ and $h$ bosons to the DM vanish,  couplings of the Standard Model to other particles of the dark sector need not.  The relic abundance can then be achieved via coannhilation \cite{Griest:1990kh} when there are other dark sector particles within 5\%-10\% of the DM mass.  In the following section, we will discuss the determination of the relic abundance in some detail.

\section{Viable Parameter Space}\label{sec:results}

In this section, we see how i) direct detection bounds and ii) the observed relic abundance constrain the singlet-doublet model. In Sec.~\ref{sec:RelicDensity}, we present points that satisfy these two conditions.  In Sec.~\ref{sec:MassSpectrum}, we extract lessons for the mass spectrum of the dark sector.  Armed with this understanding of the spectrum, in Sec.~\ref{sec:earlyUniverse}, we discuss the early Universe processes that are important for realizing the  observed dark matter density.  We then comment on implications of the spectrum for production and decays at colliders in Sec.~\ref{sec:colliders}.  

In analysis of the model, we made use of \textsc{SARAH v4.15.2}
\cite{Staub:2008uz, Staub:2013tta, Staub:2015kfa,Vicente:2015zba} to derive the interaction vertices and 2-loop RGEs.
We then employed
\textsc{SPheno v4.0.5}
\cite{Porod:2011nf,Porod:2003um}
to numerically compute the physical masses and couplings of the neutral Majorana $\chi^0_{1,2,3}$ and the charged Dirac $\chi^\pm$ fermions.
We calculated relic densities\footnote{When calculating relic densities with \textsc{micrOMEGAs}, we employ the fast calculation option (\textsc{fast=1}) and neglect the off-shell gauge bosons in the final state (\textsc{VW/VZdecay=0}).}
and direct detection cross sections using
\textsc{micrOMEGAs v6.0}
\cite{Alguero:2023zol}.
The model files for \textsc{SPheno} and \textsc{micrOMEGAs} are generated by \textsc{SARAH}.
By default, the \textsc{micrOMEGAs}
package numerically solves the Boltzmann equation and computes the relic density under the assumption that the dark sector thermalizes.  We discuss this assumption in more detail below.

With a goal of identifying parameter space that simultaneously explains the observed relic density while avoiding direct detection bounds, we performed a scan over the four-dimensional parameter space ($M_S, M_D, y_1, y_2$).  We scanned $\sim 50$ million points in the range $0.7 \lesssim M_S/M_D \lesssim 1.2$, $-1 \leq -y_2/y_1 \leq 1$, $10^{-7}\leq y_1 \leq 1.5$, and $85 \leq M_S \leq 1500$ GeV. Our focus on a region with a relatively small hierarchy between $M_{S}$ and $M_{D}$ preferentially selected out regions where coannhilation is possible.  In additional exploratory scans, we did not find points where the relic density was achieved without coannihilation.   
For each point in parameter space, we recorded the tree-level mass eigenvalues, the elements of the mass-mixing matrix, the spin-independent and spin-dependent direct detection cross sections off both protons and neutrons, the freeze-out temperature (i.e. $x_\text{FO}$), as well as the leading three processes contributing to freeze-out and the percentage of the freeze-out that they are responsible for.  Finally, we investigated the minimum reaction rate between subcomponents of the dark sector.  These data for the subset of scan points that reproduce the correct relic density are available online \cite{websiteFordata}. 

The reaction rate between dark sector subcomponents is important. For very small values of the Yukawa couplings $y_i$ chemical equilibrium between the various species within the dark sector may not be obtained.  If parts of the dark sector fall out of equilibrium, this must be taken into account.  For our analysis, we restrict ourselves to scenarios in which the dark sector remains in thermal equilibrium. Concretely, we require that the couplings are nonzero and that $\Gamma/(H x_\text{FO}) > 50$ \cite{Alguero:2023zol}.
In practice, this involves restricting results to points where at least one Yukawa coupling is larger than around $10^{-6}$.
For discussion of this model in the regime where  thermalization does not occur and the related freeze-in phenomena, see \cite{Calibbi:2018fqf}.  A closely related model is discussed in this regime in \cite{Paul:2024prs}.  

\subsection{Relic density and direct detection}\label{sec:RelicDensity}

\begin{figure}
\centering
\begin{subfigure}{0.495\textwidth}
\includegraphics[width=\textwidth]{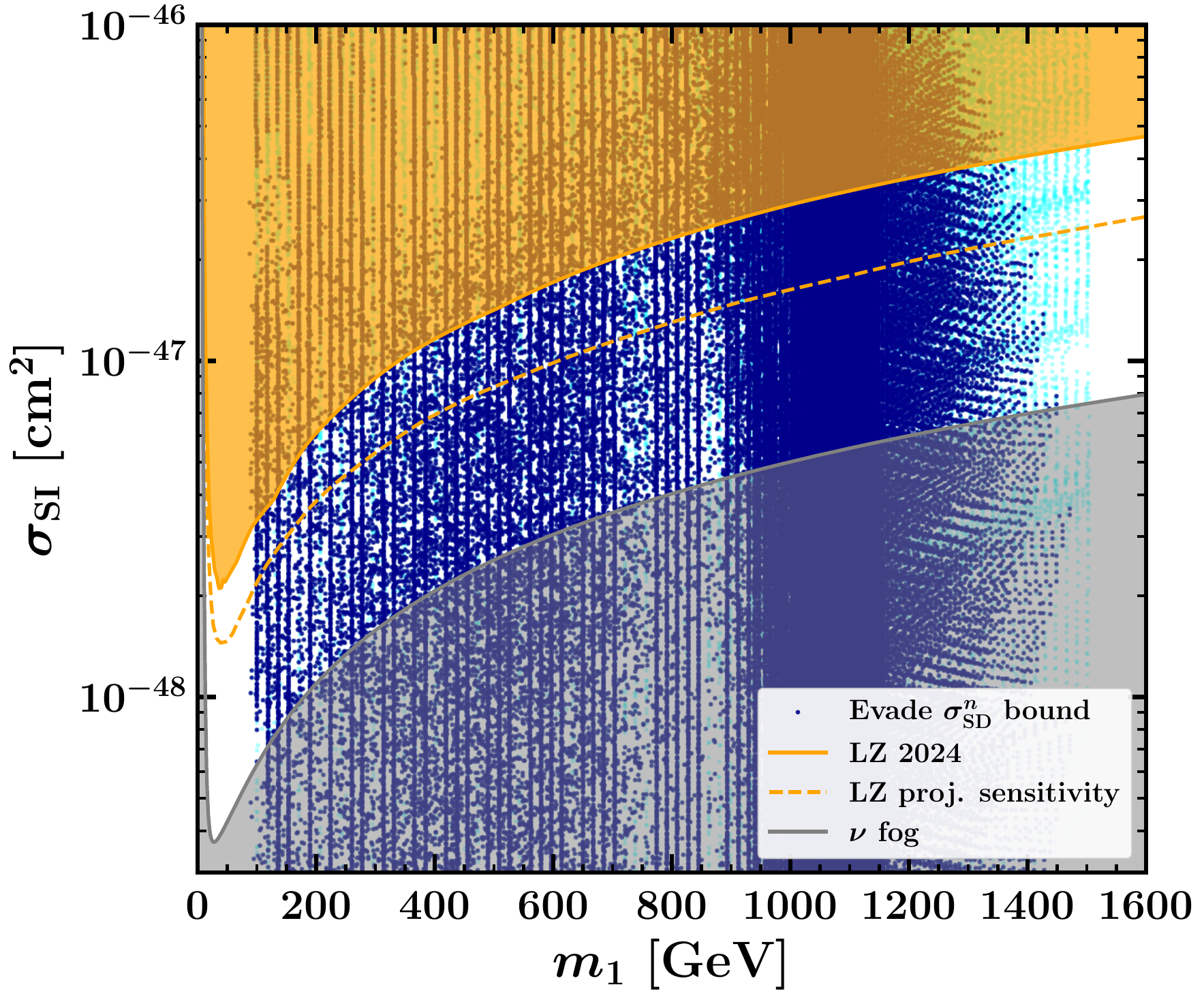}
\end{subfigure}
\begin{subfigure}{0.495\textwidth}
\includegraphics[width=\textwidth]{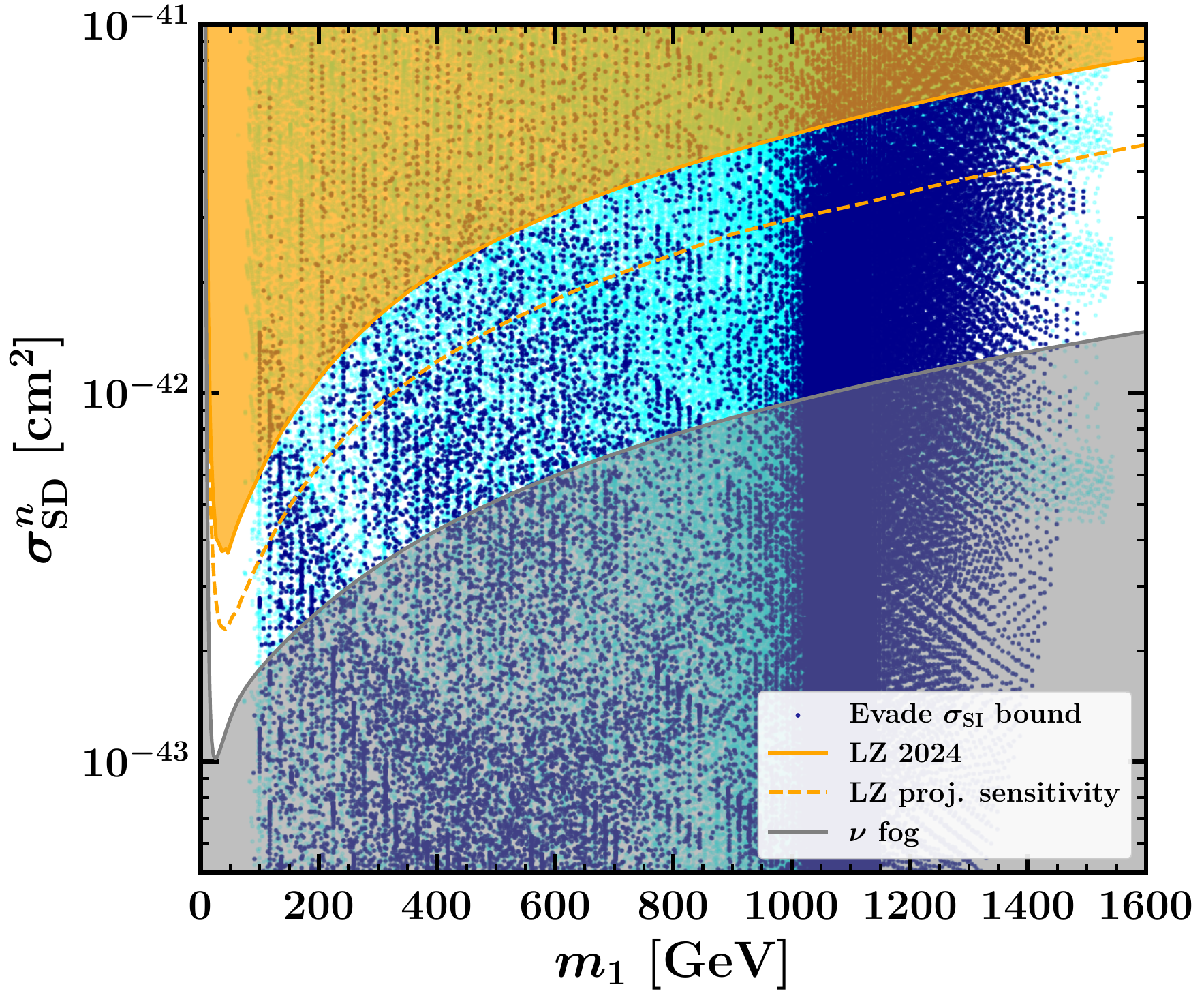}
\end{subfigure}
\caption{\textbf{Left:} The spin-independent cross section as a function of dark matter mass for parameter scan points that satisfy the observed relic density. The 90\% CL exclusion limit from the LZ collaboration in 2024 \cite{LZCollaboration:2024lux} and the projected sensitivity \cite{LZ:2018qzl} are shown in gold. The neutrino fog \cite{OHare:2021utq} is shown in gray. Light blue points satisfy the relic density while dark blue is the subset of parameter points that evade the LZ 2024 spin-dependent bound. \textbf{Right:} The spin-dependent cross section off neutrons for points satisfying the relic density.  Here the dark blue points are those that evade the LZ 2024 spin-independent bound.}
\label{fig:SIandSDXsection}
\end{figure}

In delineation of the surviving parameter space, the most important constraints on the model are that it reproduce the correct dark matter abundance and evade limits from direct detection experiments.
Figure \ref{fig:SIandSDXsection} shows the spin-independent cross section and spin-dependent cross section as a function of dark matter mass for points that satisfy the relic density constraint (light blue points).
We require $0.9\times0.12 < \Omega h^2 < 1.1\times0.12$.  The experimental measurement of the dark matter density \cite{Planck:2018vyg} is more precise than this; this range is chosen to accommodate both the experimental and theoretical uncertainties involved in the calculation.
Shown are the 90\% CL exclusion limits from the LZ collaboration in 2024 \cite{LZCollaboration:2024lux},
along with the neutrino fog from Ref.~\cite{OHare:2021utq}.
Projections and previous results from the LZ collaboration are found in Refs. \cite{LZ:2018qzl, LZ:2022lsv}.  Shown in dark blue on the spin-independent (spin-dependent) plot are the parameter space points that evade the spin-dependent (spin-independent) bound.
Although many points that evade one bound are constrained by the other, the spin-independent and spin-dependent direct detection cross sections can both be small.  Indeed, the existence of dark blue points beneath the neutrino fog indicates that points  will remain stubbornly out of reach of direct detection going forward.  We note we have intentionally explored parameter space that can simultaneously achieve the relic density while having allowed direct detection cross sections.  Had we not done so, and had, for example, uniformly sampled from a larger parameter space, the vast majority of points would be excluded by the direct detection bounds.

We have established small direct detection cross sections are consistent with the observed DM density. We now focus on characterizing these points, i.e. describing the parameter space that both evades the 2024 LZ  constraints \cite{LZCollaboration:2024lux} and reproduces the correct DM abundance. 

\begin{figure}
    \centering
    \includegraphics[width=0.8\textwidth]{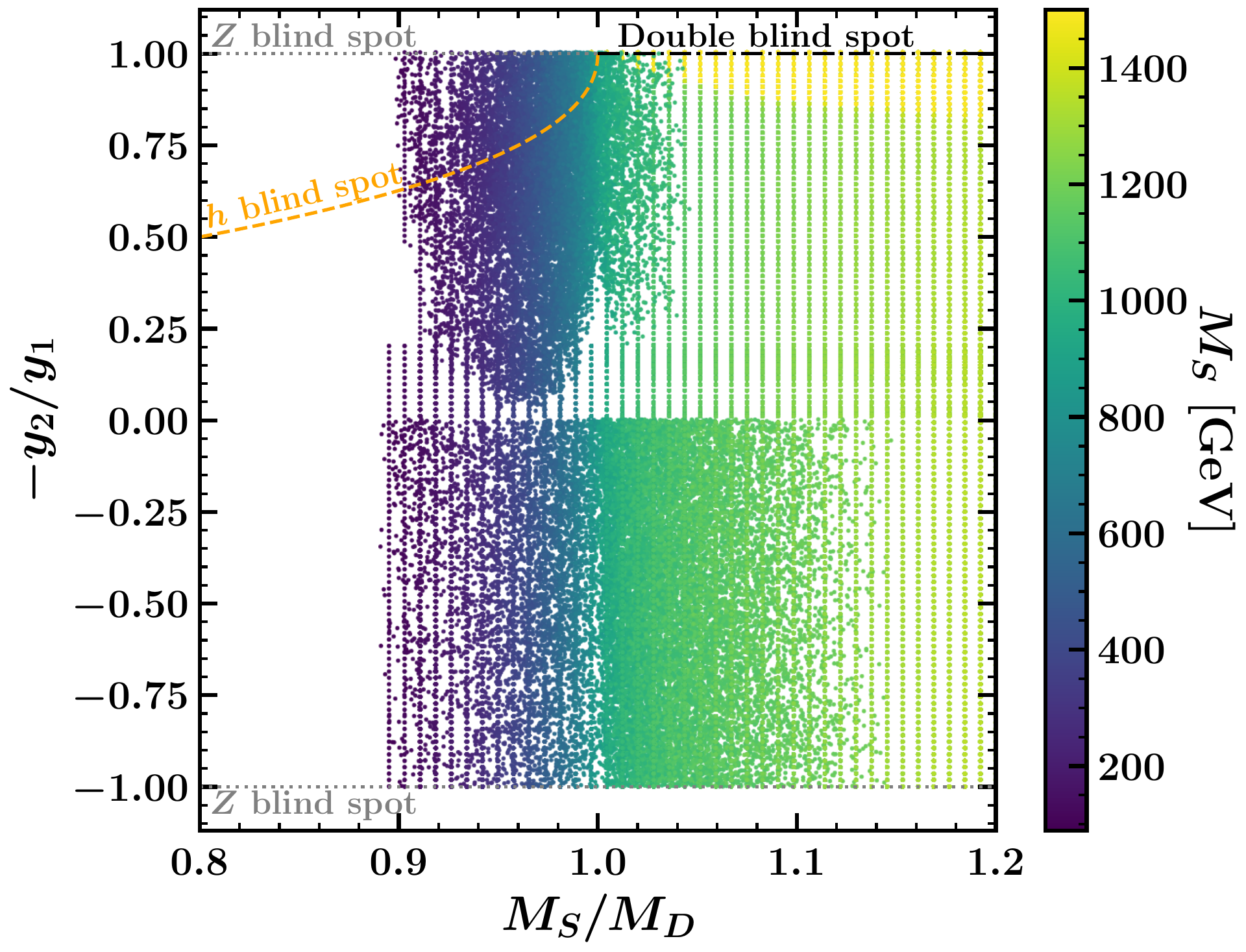}
    \caption{Coupling ratio $-y_2/y_1$ vs mass ratio $M_S/M_D$ for parameter scan points that evade the LZ 2024 direct detection constraints \cite{LZCollaboration:2024lux} and reproduce the correct relic density.  The color bar indicates $M_S$ values for these points.
    The  $Z$, $h$, and double blind spots, corresponding to Eqs. (\ref{eq:ZBS})-(\ref{eq:DoubleBS}), are shown as dotted, dashed, and dash-dotted lines, respectively.}
    \label{fig:CoupRat_Vs_MassRat_MsCbar}
\end{figure}

Figure \ref{fig:CoupRat_Vs_MassRat_MsCbar} shows where in the $-y_2/y_1$ vs $M_S/M_D$ plane these viable 
points lie.  The color bar shows the $M_S$ values for each of these points.  The relation between the Yukawa couplings that satisfy the Higgs blind spot condition (Eq.~(\ref{eq:HBS})) is shown as a dashed line, whereas the $Z$ blind spots (Eq.~(\ref{eq:ZBS})) are indicated as dotted lines.
The Higgs blind spot and one of the $Z$ blind spots ($-y_2/y_1 = 1$)
coincide for $M_{D} < M_{S}$, where we have shown them as a dash-dotted line (Eq.~(\ref{eq:DoubleBS})). For $M_S \lesssim 850$ GeV, we see that it is possible to have $M_S/M_D \lesssim 1$.  As $M_S$ increases, the mass ratio $M_S/M_{D}$ increases, and for the highest masses, points tend to cluster near the double blind spot (i.e. $-y_2/y_1 \sim 1$).  However, in this projection, many points also appear far from the blind spot(s).

\begin{figure}
\centering
\begin{subfigure}{0.5\textwidth}  
\includegraphics[width=\textwidth]{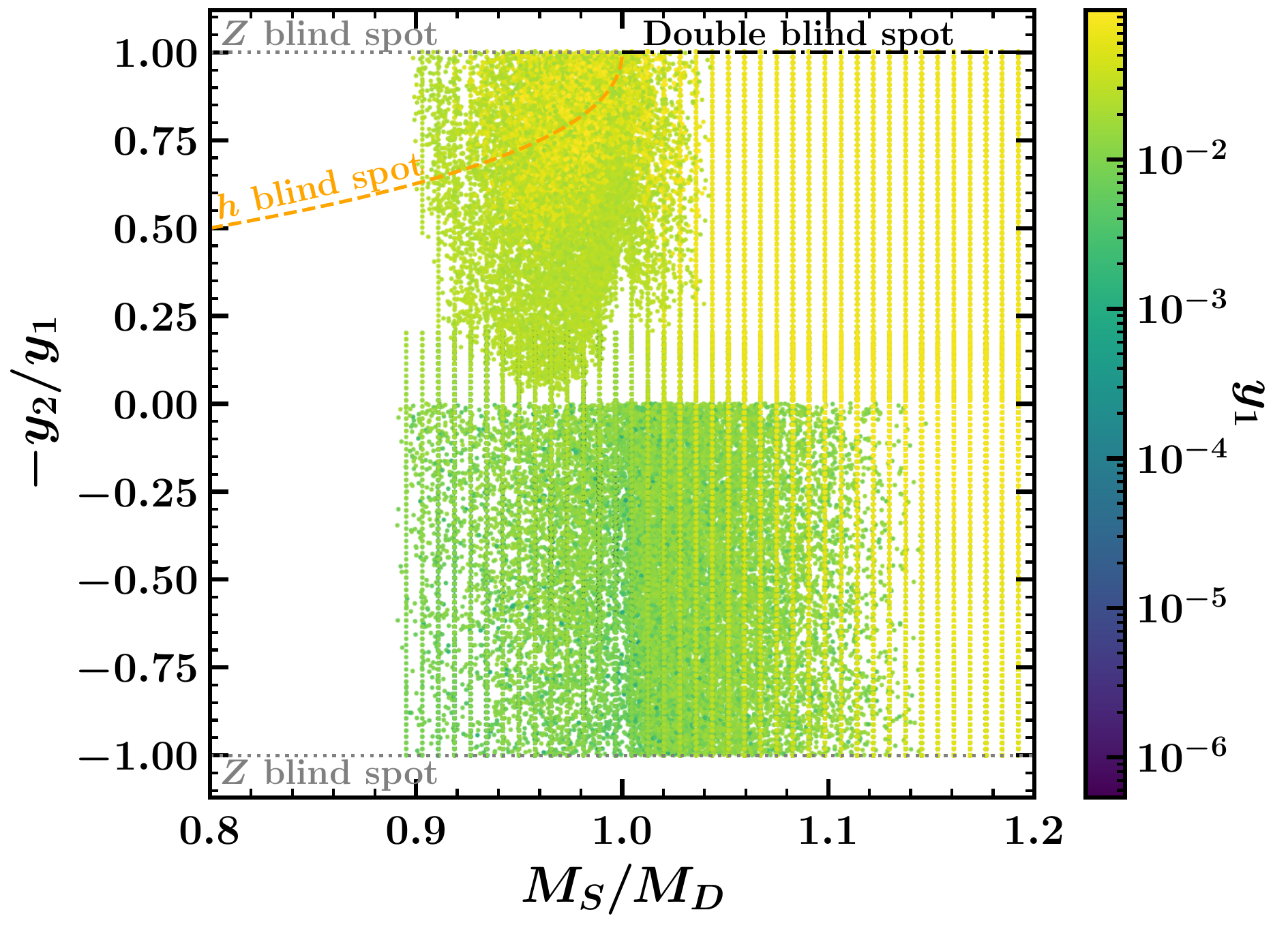}
\end{subfigure}
\begin{subfigure}{0.49\textwidth}  
\includegraphics[width=\textwidth]{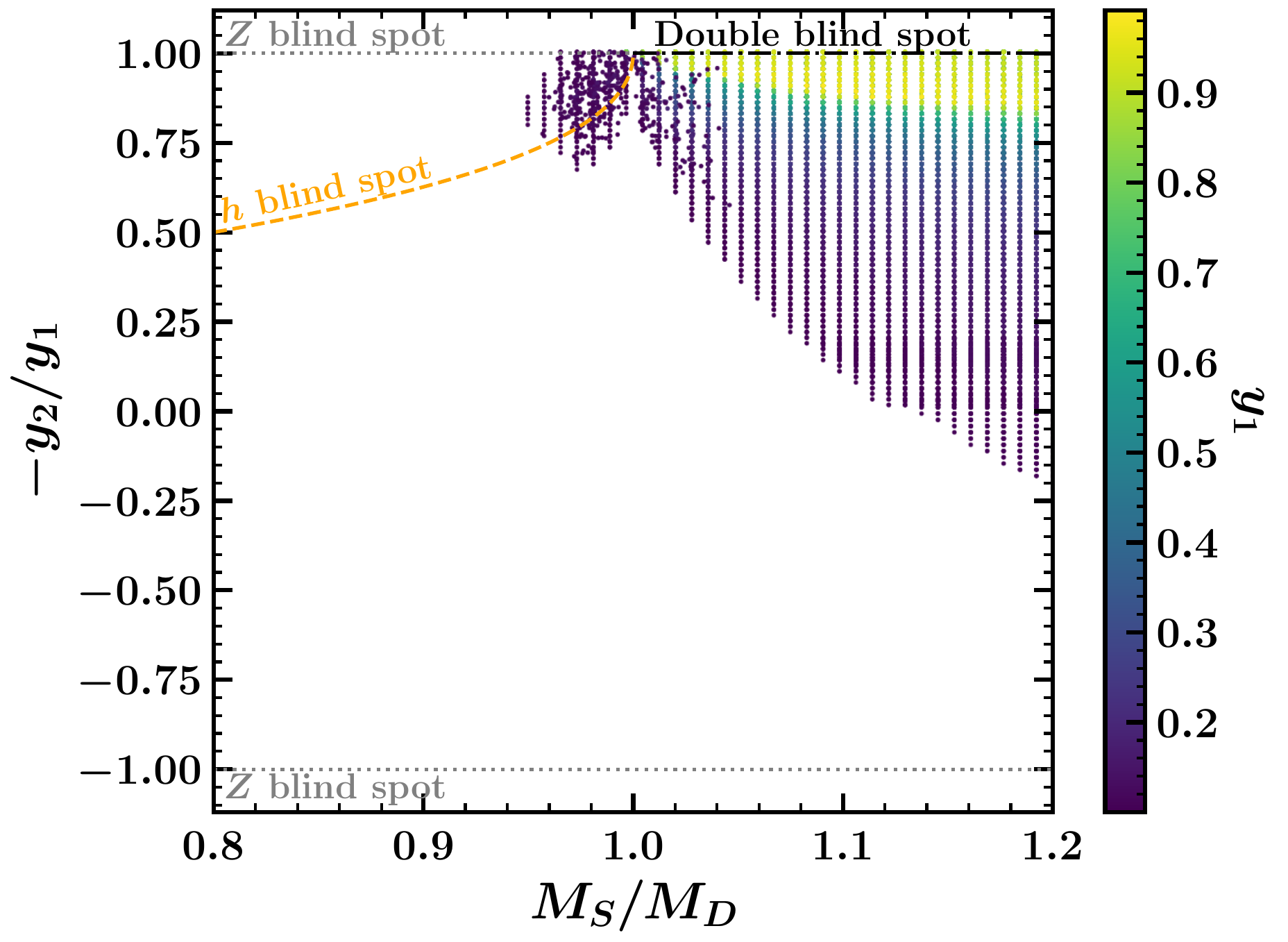}
\end{subfigure}
\caption{Coupling ratio $-y_2/y_1$ vs mass ratio $M_S/M_D$ color coded by $y_1$ values for all points that evade the LZ 2024 direct detection constraints \cite{LZCollaboration:2024lux} and reproduce the correct relic density.
The blind spots, defined in Eqs.~(\ref{eq:ZBS})-(\ref{eq:DoubleBS}), are shown as labeled.
\textbf{Left:}  In this plot, we restrict to $y_1 < 0.1$. We see that for small $y_1$ values all values of $-y_2/y_1$ are allowed. \textbf{Right:} In this plot, we restrict to $y_1 \geq 0.1$.  For larger the $y_1$,  $-y_{2}/y_{1}$ must be closer to blind spots for consistency with direct detection constraints.} 
\label{fig:CoupRat_Vs_MassRat_SmallandLargeY1s}
\end{figure}

To elucidate what is happening for these points, in
Figure \ref{fig:CoupRat_Vs_MassRat_SmallandLargeY1s} we again show surviving points in the $-y_2/y_1$ vs $M_S/M_D$ plane but now color-coded by $y_1$ value. In the left panel we show the smallest $y_1$ values ($ y_1 < 0.1$); the right panel contains the remaining $y_1$ values. For small $y_1$, it is possible to inhabit a wider range of parameter space.  But for $y_1 \gtrsim 0.1$, $y_1$ and $y_2$ are required to be of opposite sign (i.e. $-y_2/y_1 \gtrsim 0$). The choice $-y_2/y_1 \gtrsim 0$ places the model in the half-plane that contains both spin-independent and spin-dependent blind spots, and couplings are generically smaller in this region. 

For even larger $y_1$ values,  $-y_2/y_1 \gtrsim 0$ is insufficient to avoid direct detection constraints.  Parameters must be even closer to blind spots.  For $M_S < M_D$, since the spin-independent and spin-dependent blind spots do not coincide, we find that the maximum $y_1$ that can evade both direct detection constraints is $y_1 \sim 0.2$. 
In the MSSM, the analogous values of $y_{1}^2 + y_{2}^2$ are fixed to be proportional to the hypercharge gauge coupling. So, in the MSSM this constraint on $y_{1}$ may be reinterpreted as a constraint on $\tan \beta$ in the regime where $M_1 < \mu$. Only $\tan \beta ~\sim 1$ are allowed, which would be consistent with the observed value of the Higgs boson mass only for heavy scalar masses, i.e. split supersymmetry.

As $M_{S}\rightarrow M_{D}$ the blind spot  suppresses couplings to both the $Z$ and Higgs bosons.  As a result, direct detection signals are suppressed and $y_1 \sim 1$ is allowed for points in close proximity to this double blind spot.
As $M_S$ continues to increase past $M_D$, values of $-y_{2}/y_{1}$ that depart further from the blind spot are allowed. There are two reasons for this.  First, as we move into this region, the mass scale of the dark matter increases and the direct detection limits weaken linearly with increasing mass. Second, as $M_S - M_D$ increases, mixing between the two states decreases for a given choice of Yukawa couplings (see Appendix \ref{sec:appSmally}).  This means that the dark matter can be closer to the pure doublet DM limit, for which direct detection is suppressed, even in the presence of larger Yukawa couplings. 

In the $M_S > M_D$ region of parameter space, there can be multiple states that are close in mass.  We now comment on the case where the lightest two neutral states are \emph{very} nearly degenerate.  In this case, the dark matter may scatter inelastically into the companion state, potentially with a large rate, mediated by a vector coupling to the $Z$ boson.   That is, the dark matter becomes pseudo-Dirac in nature, and in the limit of exact degeneracy, the value of the direct detection cross section would approach that of a Dirac neutrino, several orders of magnitude in excess of current bounds.  To avoid this possibility, we impose that the mass difference between the two lightest neutral states be greater than $200$ keV.  This provides an effective kinematic barrier, due to the finite escape velocity in our galaxy \cite{Tucker-Smith:2001myb, Graham:2024syw}.  For fixed values of the Yukawa coupling, this effectively places an upper bound on the value of $M_{S}$. Also, in the limit of very small mass differences, both states inhabit the bath at freeze-out, and the $\chi^0_2$ eventually decays as $\chi^0_2 \rightarrow \chi^0_1 \gamma$, adding to the dark matter density.  After imposing the minimum splitting discussed above, these decays occur before big bang nucleosynthesis (BBN), and so will not run afoul of cosmological constraints.

For both Figure \ref{fig:CoupRat_Vs_MassRat_MsCbar} and Figure \ref{fig:CoupRat_Vs_MassRat_SmallandLargeY1s} the precise shape of the allowed regions are influenced by the chosen scanning strategy.   However, that $M_S/M_D$ $\gtrsim$ .9 for all points shown is significant -- this indicates the necessity for coannhilation to achieve the DM relic density.
 And while when $M_S < M_D$,  effective coannihilation requires  $M_S \sim M_D$, when $M_D < M_S$,  other states (i.e. the rest of the doublet) are automatically close in mass to the dark matter independent of whether $M_S$ is close to $M_{D}$.  Coannihilation, therefore, does not impose a maximum value of $M_S/M_D$ in Figures \ref{fig:CoupRat_Vs_MassRat_MsCbar} and  \ref{fig:CoupRat_Vs_MassRat_SmallandLargeY1s}. 

\begin{figure}
\centering

\begin{subfigure}{0.328\textwidth}
\includegraphics[width=\textwidth]{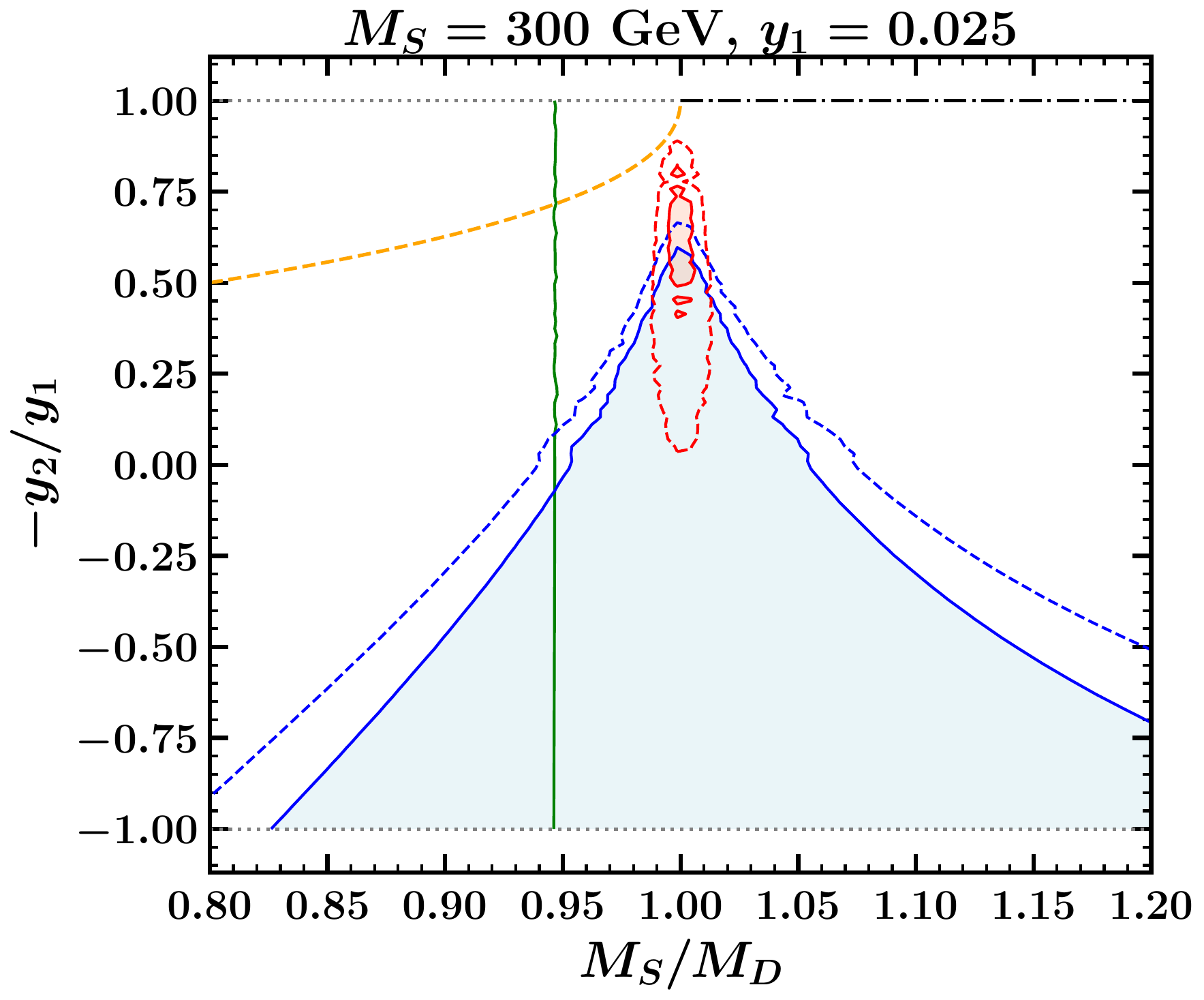}
\end{subfigure}
\begin{subfigure}{0.328\textwidth}
\includegraphics[width=\textwidth]{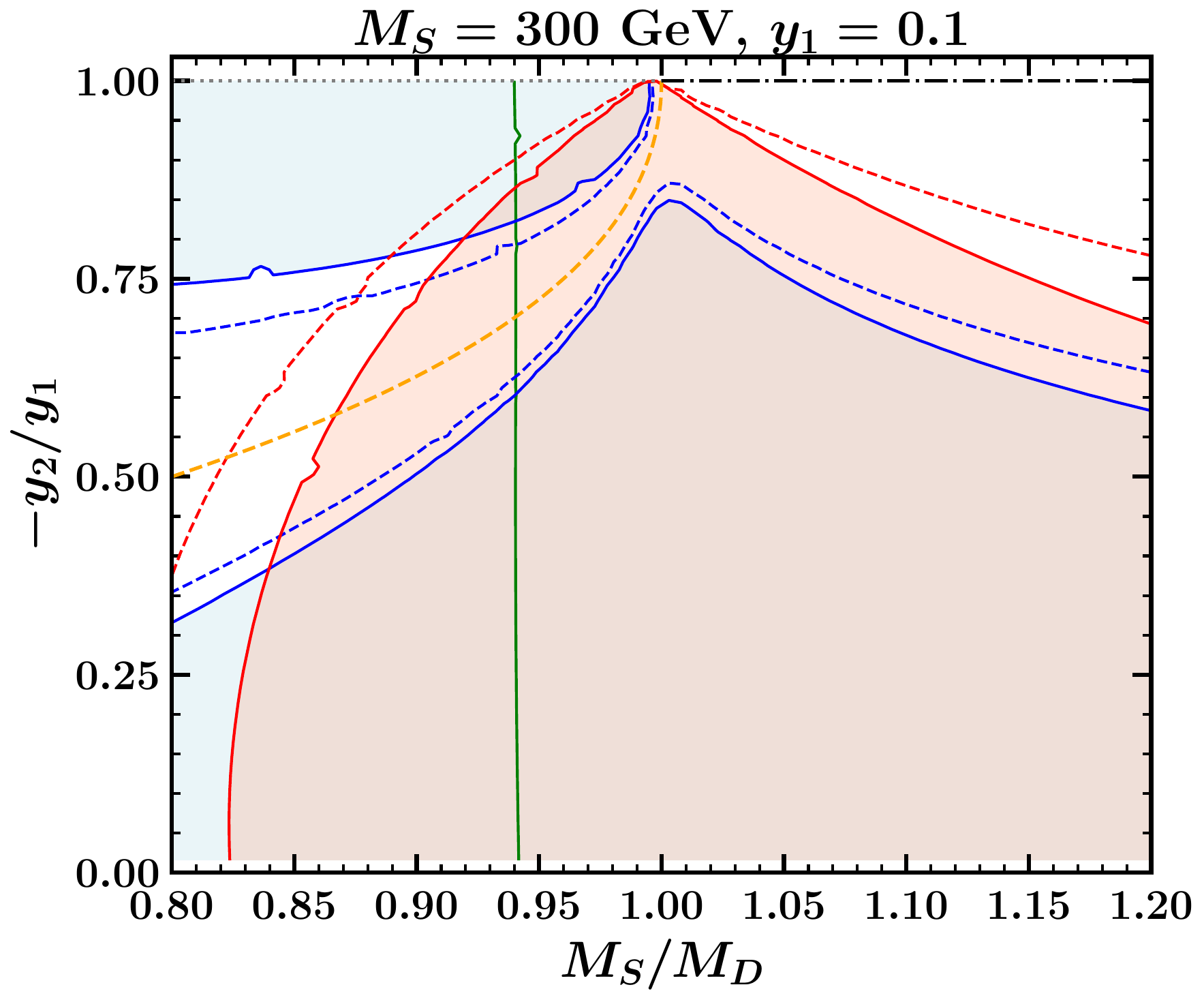}
\end{subfigure}
\begin{subfigure}{0.328\textwidth}
\includegraphics[width=\textwidth]{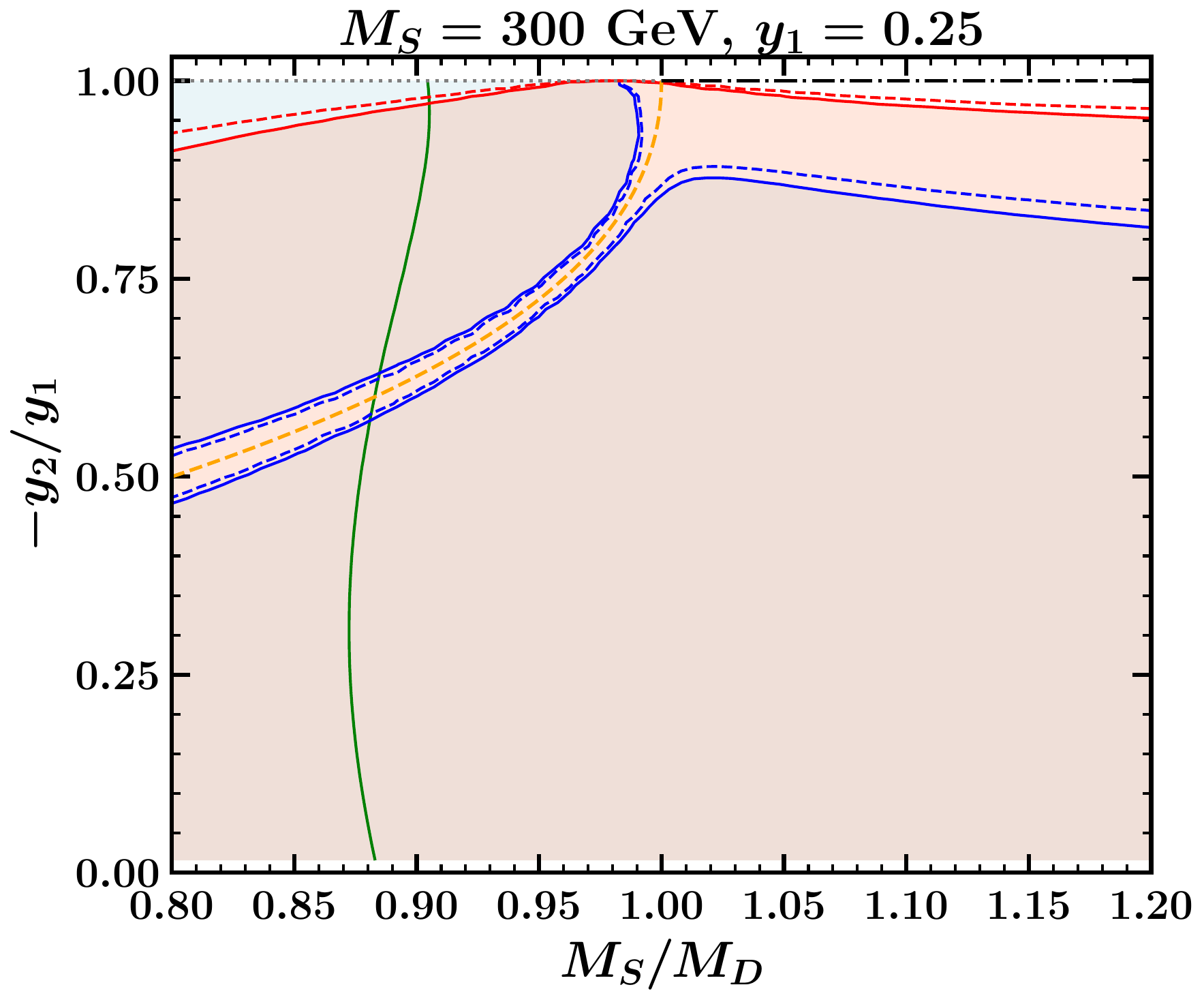}
\end{subfigure}

\begin{subfigure}{0.328\textwidth}
\includegraphics[width=\textwidth]{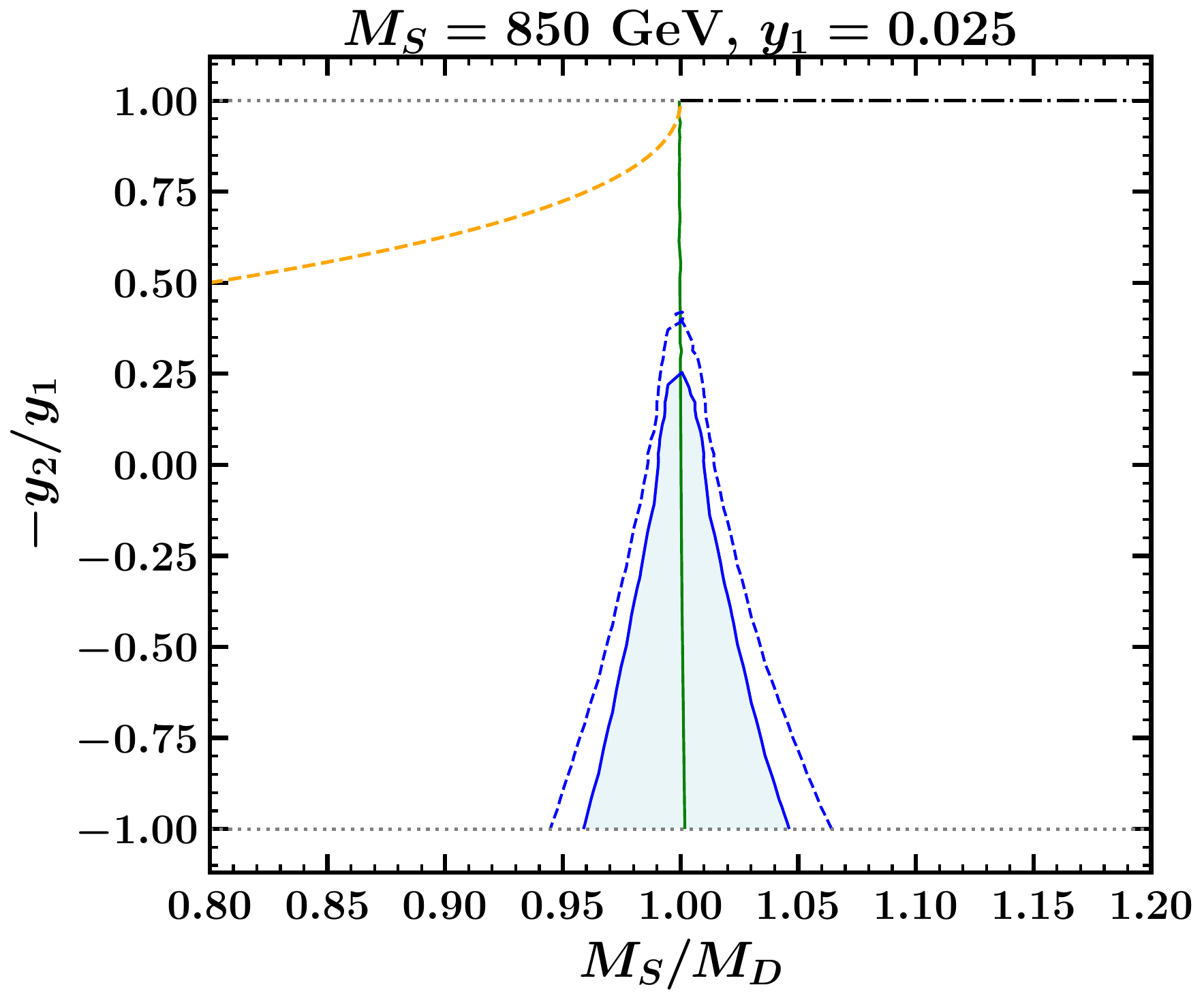}
\end{subfigure}
\begin{subfigure}{0.328\textwidth}
\includegraphics[width=\textwidth]{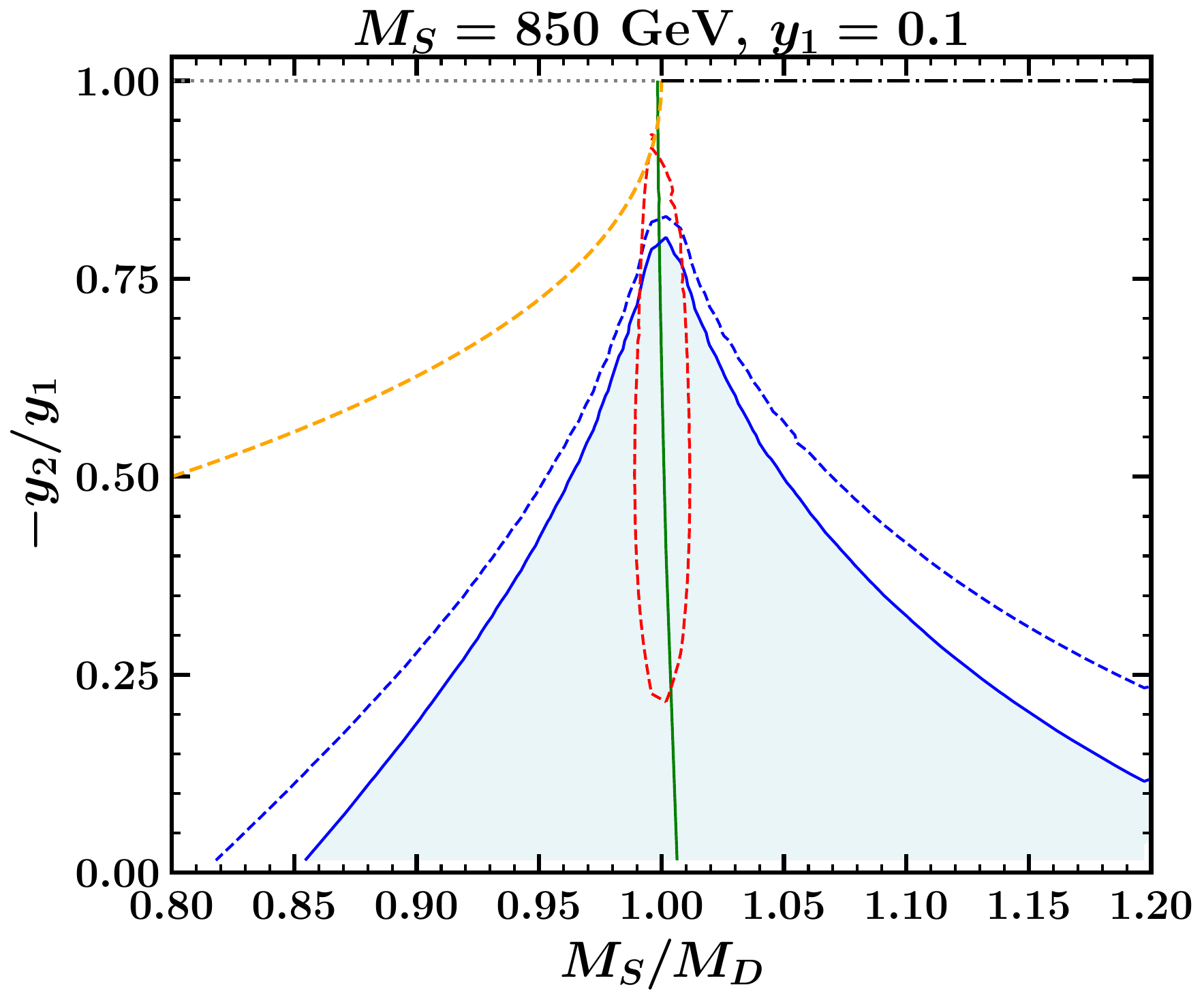}
\end{subfigure}
\begin{subfigure}{0.328\textwidth}
\includegraphics[width=\textwidth]{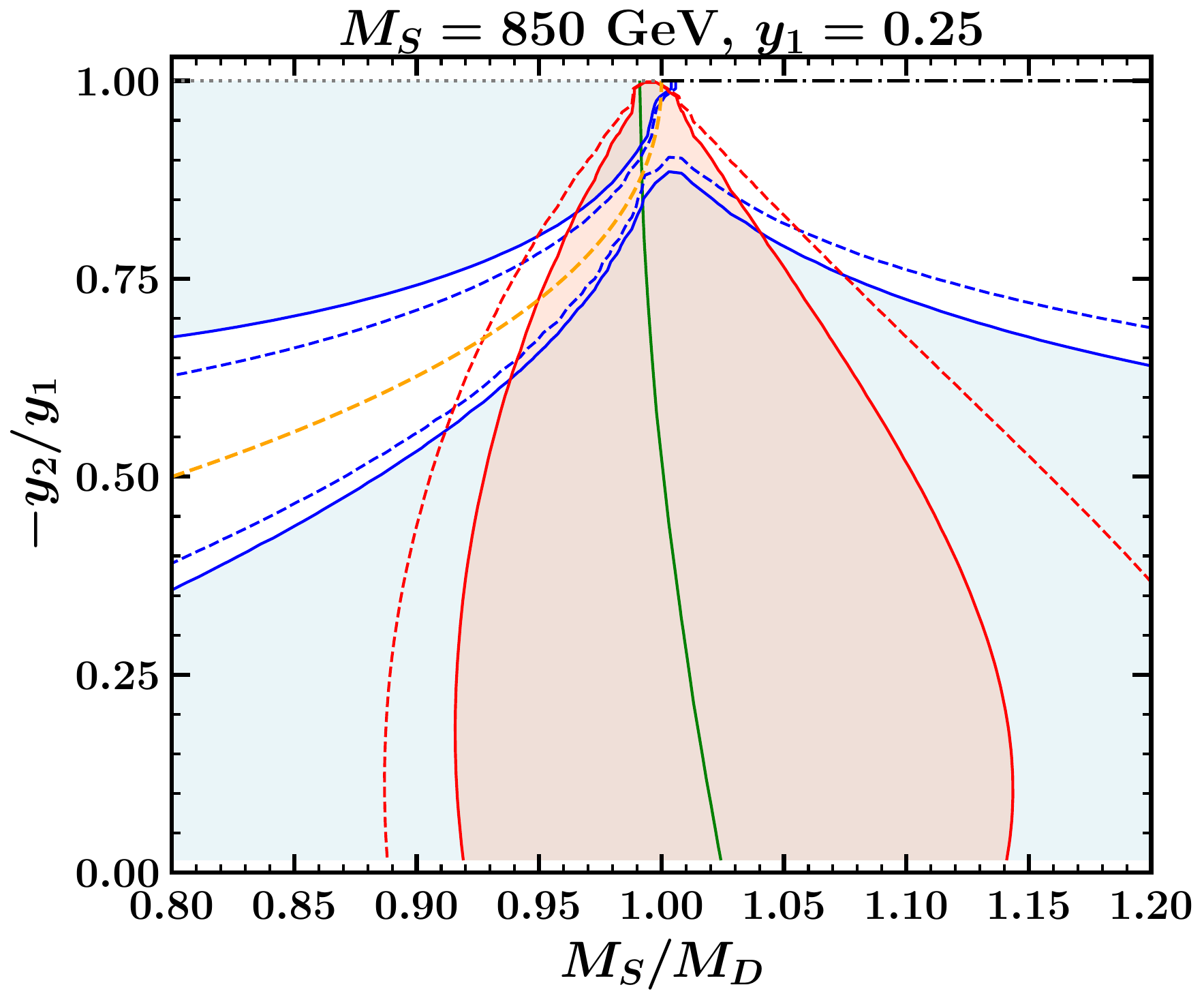}
\end{subfigure}

\begin{subfigure}{0.328\textwidth}
\includegraphics[width=\textwidth]{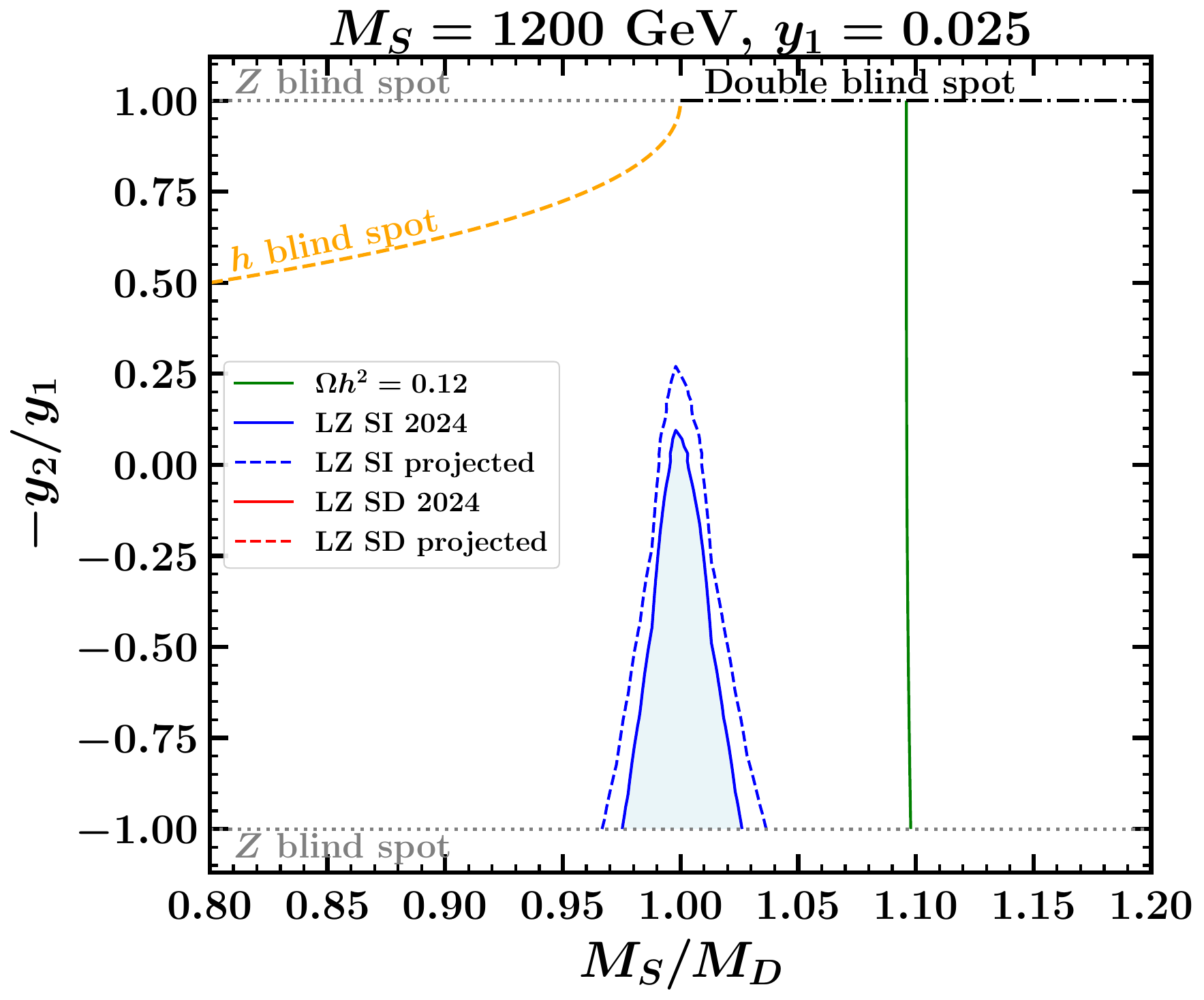}
\end{subfigure}
\begin{subfigure}{0.328\textwidth}
\includegraphics[width=\textwidth]{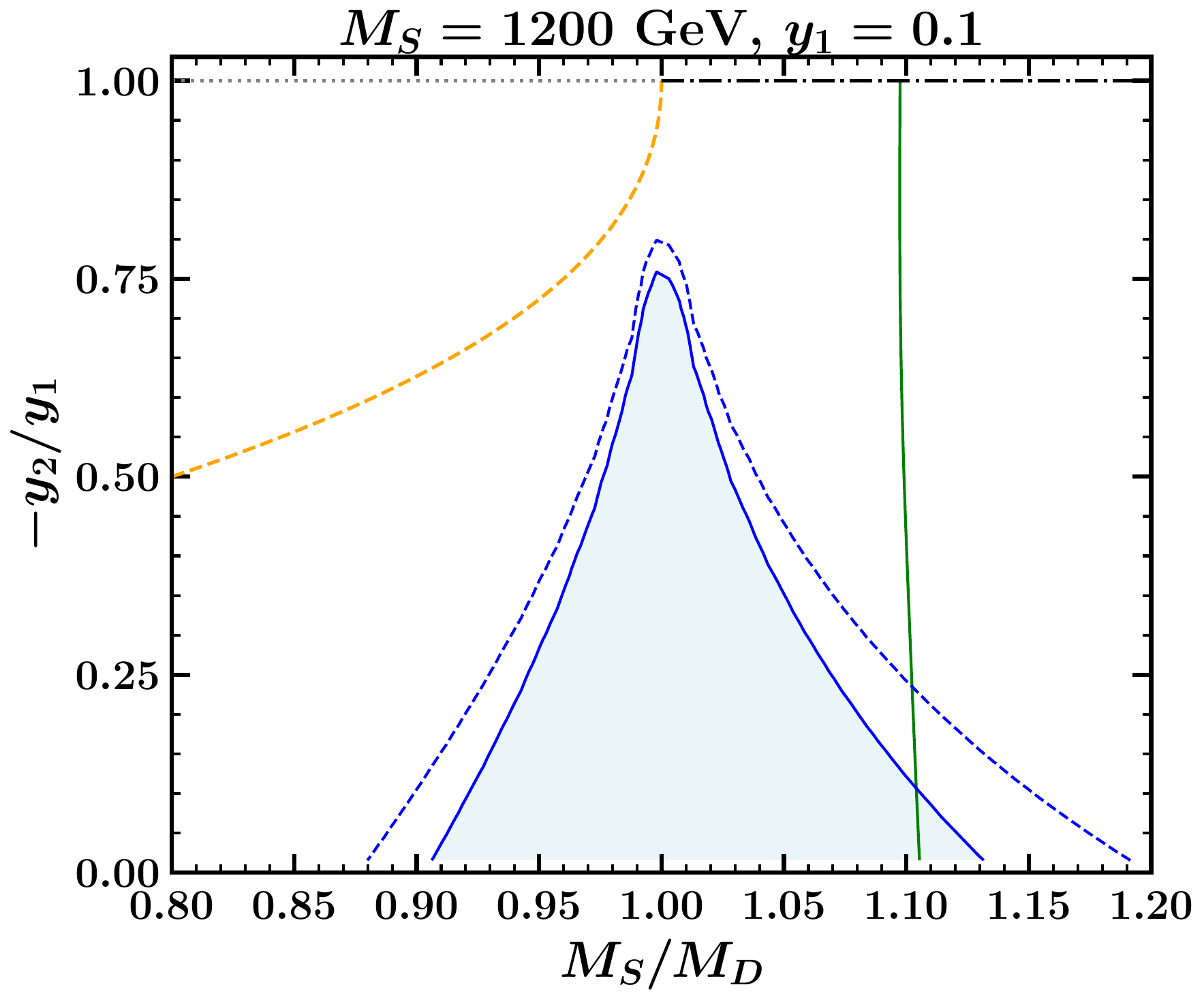}
\end{subfigure}
\begin{subfigure}{0.328\textwidth}
\includegraphics[width=\textwidth]{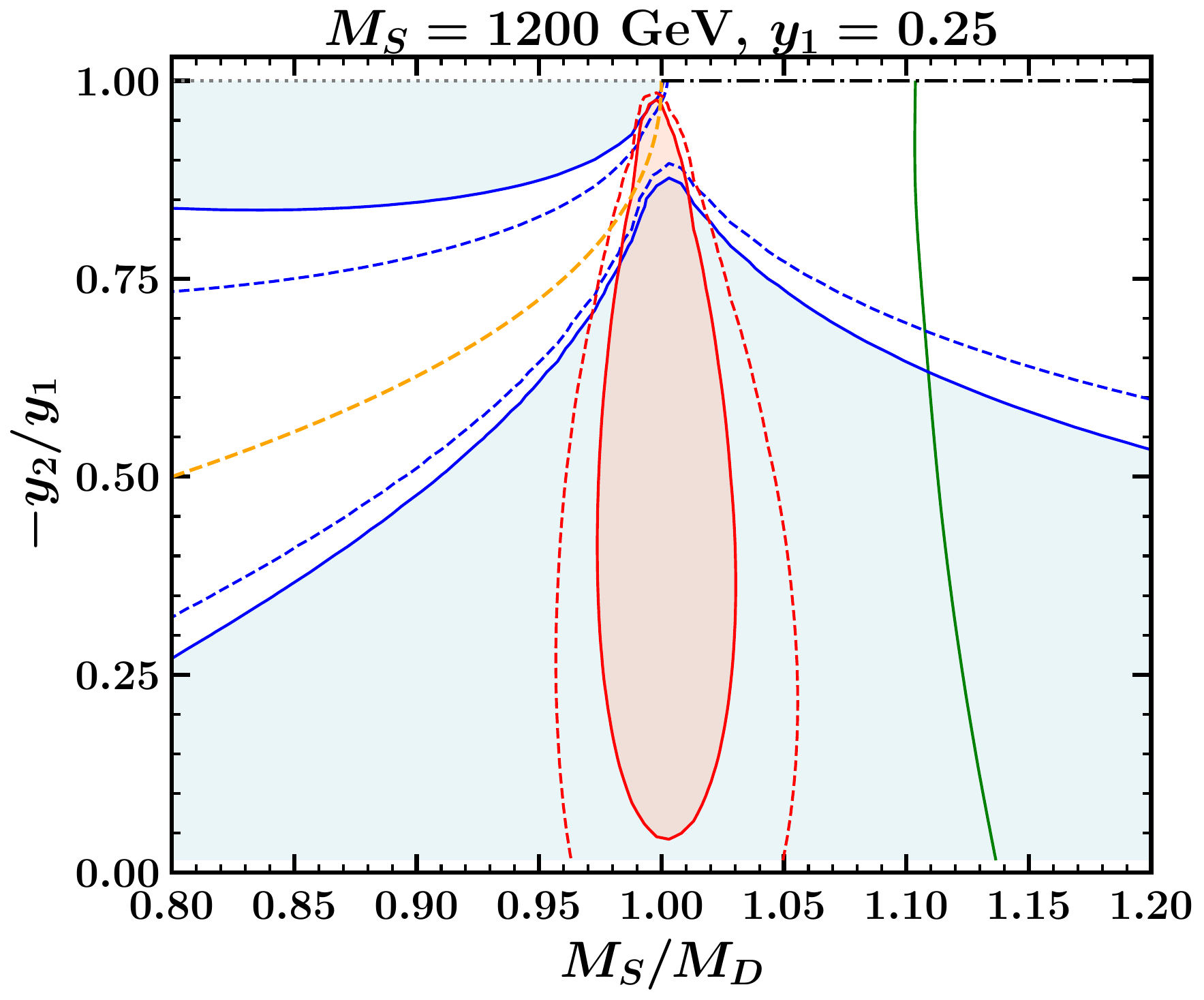}
\end{subfigure}
\caption{Direct detection bounds and relic density curves for different choices of $M_S$ and $y_1$.  Areas currently excluded by the LZ 2024 spin-independent (spin-dependent) direct detection bounds \cite{LZCollaboration:2024lux} are shaded in light blue (light red).  The dashed curves of the same colors proximal to these regions indicate how these regions will expand when LZ reaches its projected sensitivity \cite{LZ:2018qzl}. Points that satisfy the relic density constraint lie along the green lines.  The spin-independent (Higgs) blind spot for $M_S < M_D$ is shown as a gold dashed line. The spin-dependent ($Z$) blind spot for $M_S < M_D$ is shown as a gray dotted line.  The double blind spot for $M_D < M_S$ is drawn as a black dash-dotted line.  Points that simultaneously satisfy the relic density while avoiding direct detection occur when the green line passes through a white region. As discussed in the text, simultaneously achieving the right relic density and avoiding direct detection constraints for $y_1 \gtrsim 0.1$  requires $y_1$ and $y_2$ of opposite sign, so the y-axis range is smaller for the second and third column.}
\label{fig:ContourPlots}
\end{figure}

In Figure \ref{fig:ContourPlots} we display how direct detection limits change in the $-y_2/y_1$ vs $M_S/M_D$ plane as $M_S$ and $y_1$ are varied.  Because of the precision of the dark matter density measurement, points that realize the correct relic density appear as a curve in this plane.  Also shown is how this curve varies as $M_S$ and $y_1$ are varied.   We choose $M_S = 300, 850, 1200$ (top to bottom)
and
$y_1 = 0.025, 0.1, 0.25$ (left to right)
as benchmark values. Since there is never any relevant parameter space when $-y_2/y_1 \lesssim 0$ and $y_1 \gtrsim 0.1$ we restrict the figures to $-y_2/y_1 \geq 0$ for the larger two $y_1$ values.  

The light blue (red) shaded regions correspond to regions excluded by the LZ 2024 spin-independent (dependent) bounds, respectively. Should LZ bounds improve to their projected sensitivity, these regions will expand to the dashed curves of the same color. The green curves denote points that realize the observed dark matter relic abundance.  The $h$-blind spot is indicated with a gold-dashed line; the $Z$ blind spot for $M_{S} < M_{D}$ is shown as a gray-dotted line, and the double blind spot (for $M_D < M_S$) is indicated by a black dash-dotted line. Consistency with both the observed relic abundance and direct detection bounds requires the green line to transit the unshaded space.  For example, no such points exist in the upper-right panel ($M_S = 300$ GeV, $y_1 = 0.25$), but the entirety of the green line is allowed in the lower-left panel ($M_S = 1200$ GeV, $y_1 = 0.025$).
  
We note the following general trends.  For a fixed coupling strength $y_1$, higher $M_S$ requires a larger $M_S/M_D$ value to achieve the observed relic density.  This results in the DM becoming more  doubletlike.  Also, for fixed $y_1$, raising $M_S$ relaxes both the spin-independent and spin-dependent limits.  For fixed $M_S$,  direct detection limits require parameters that more closely satisfy the blind spot condition(s) as  $y_1$ increases.  Conversely, for the smallest $y_1$ values we have shown, direct detection limits are fairly weak and most of the $-y_2/y_1$ vs $M_S/M_D$ plane is allowed, i.e the parameters need not be close to a blind spot.  

It is notable that the green curves are insensitive to the value of $-y_2/y_{1}$, especially in the regions allowed by both the relic density and the direct detection bounds.  This indicates that neither interactions with the Higgs boson nor the precise value of the mixing between the singlet and the doublets plays a large role in the determination of the relic density.  Similarly, for a fixed value of $M_S$, the $x$ intercept of the green line is essentially unchanged for the two smallest values of the $y_{1}$ shown. This also indicates the relative insensitivity of the relic abundance to these values of Yukawa coupling.  We comment more on these points in Sec.~\ref{sec:earlyUniverse}.

\subsection{Mass spectrum} \label{sec:MassSpectrum}

\begin{figure}
    \centering
    \includegraphics[width=0.8\textwidth]{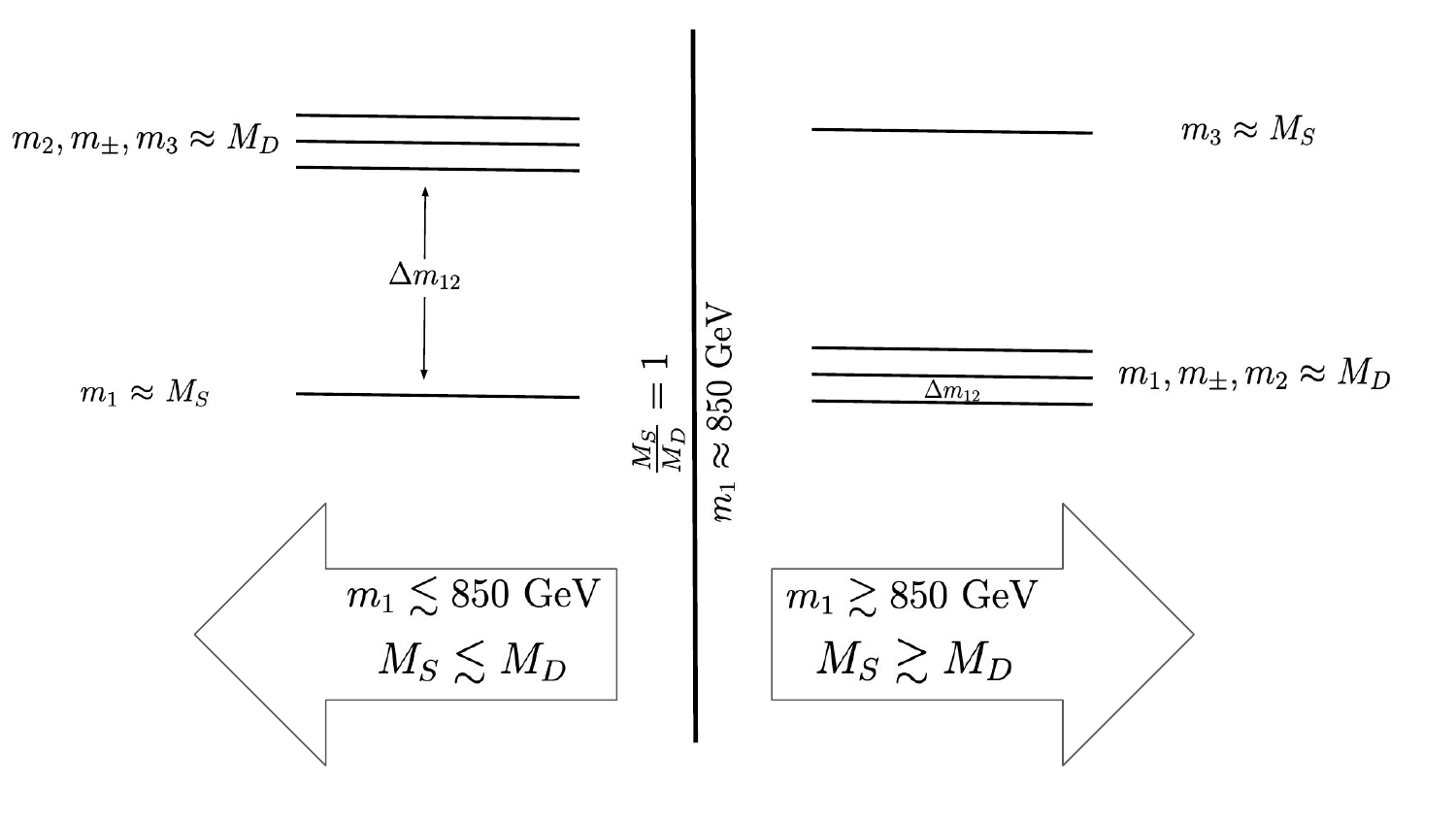}
    \caption{Qualitative features of the mass spectrum above and below $m_1 \approx 850$ GeV ($M_S = M_D$).  Below this point, the dark matter mass is approximately equal to the $M_S$ parameter and there are three states near $M_D$.  The mass difference $\Delta m_{12} \equiv m_2 - m_1$ is $\sim 1-20$ GeV.  For $M_S > M_D$, the dark matter and two other dark sector states are approximately equal to the $M_D$ parameter and $\Delta m_{12}$ can be less than a GeV.  The heaviest neutral state can be much heavier.  There is a smooth transition from the left to right of this figure as $M_S/M_D$ increases.}
    \label{fig:MassSpectSchematic}
\end{figure}

In this Section, we examine the dark sector mass spectrum for the surviving parameter space.  Understanding this spectrum aids in comprehending both the early universe annihilation mechanism and the present-day collider signatures.  For our viable parameter space, the mass eigenvalues written in terms of the mass parameters are $m_1 \approx \text{min}(M_S, M_D)$, $m_2 \approx M_D$, $m_{\pm}\approx M_D$, $m_3 \approx \text{max}(M_S, M_D)$.  Regardless of whether $M_{S}$ or $M_{D}$ is larger, there is one state near $M_S$ and three states near $M_D$. The qualitative features of the mass spectrum in the two regions are captured in Figure \ref{fig:MassSpectSchematic}.

\begin{figure}
\centering
\begin{subfigure}{0.48\textwidth}
\includegraphics[width=\textwidth]{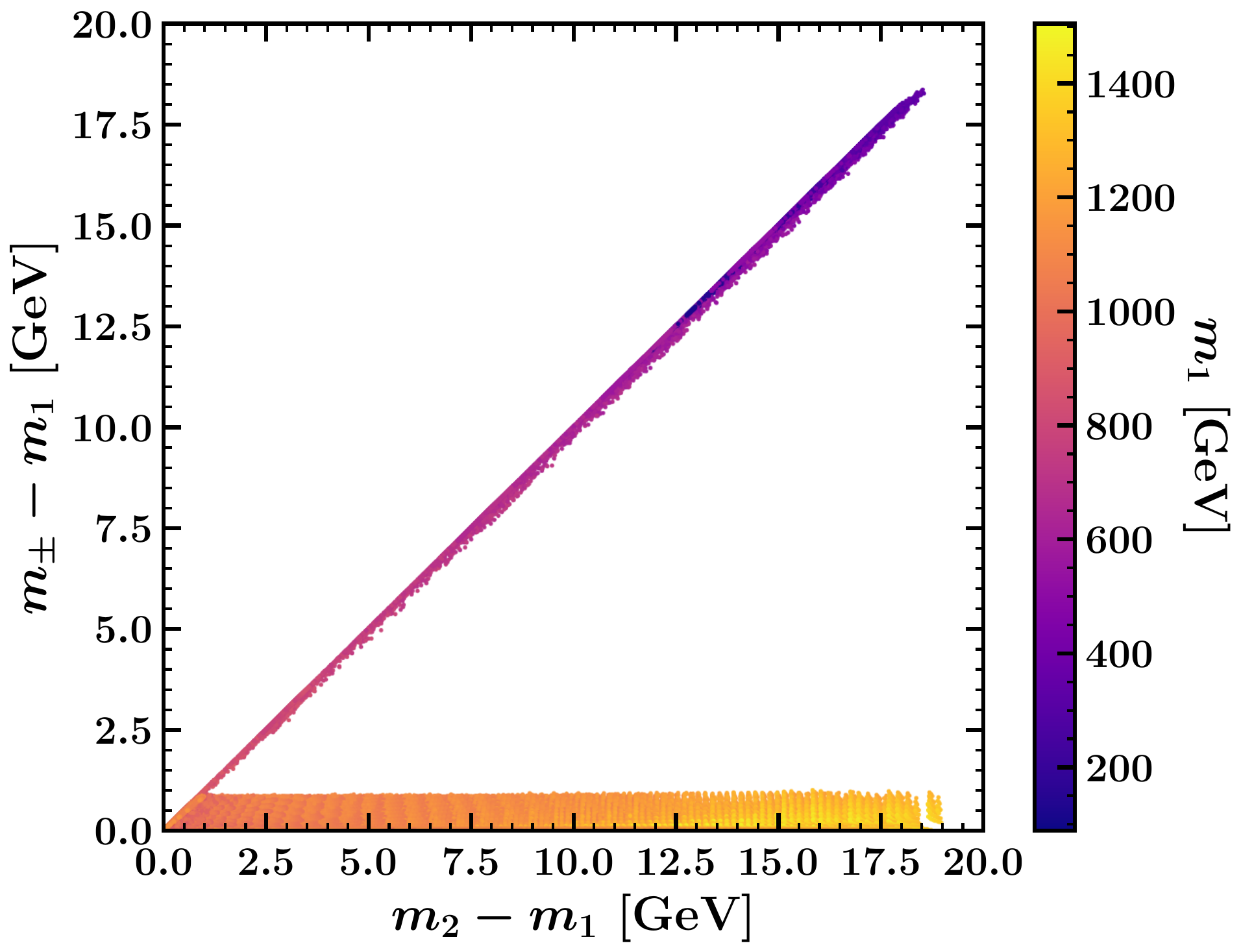}    
\end{subfigure}
\begin{subfigure}{0.48\textwidth}
\includegraphics[width=\textwidth]{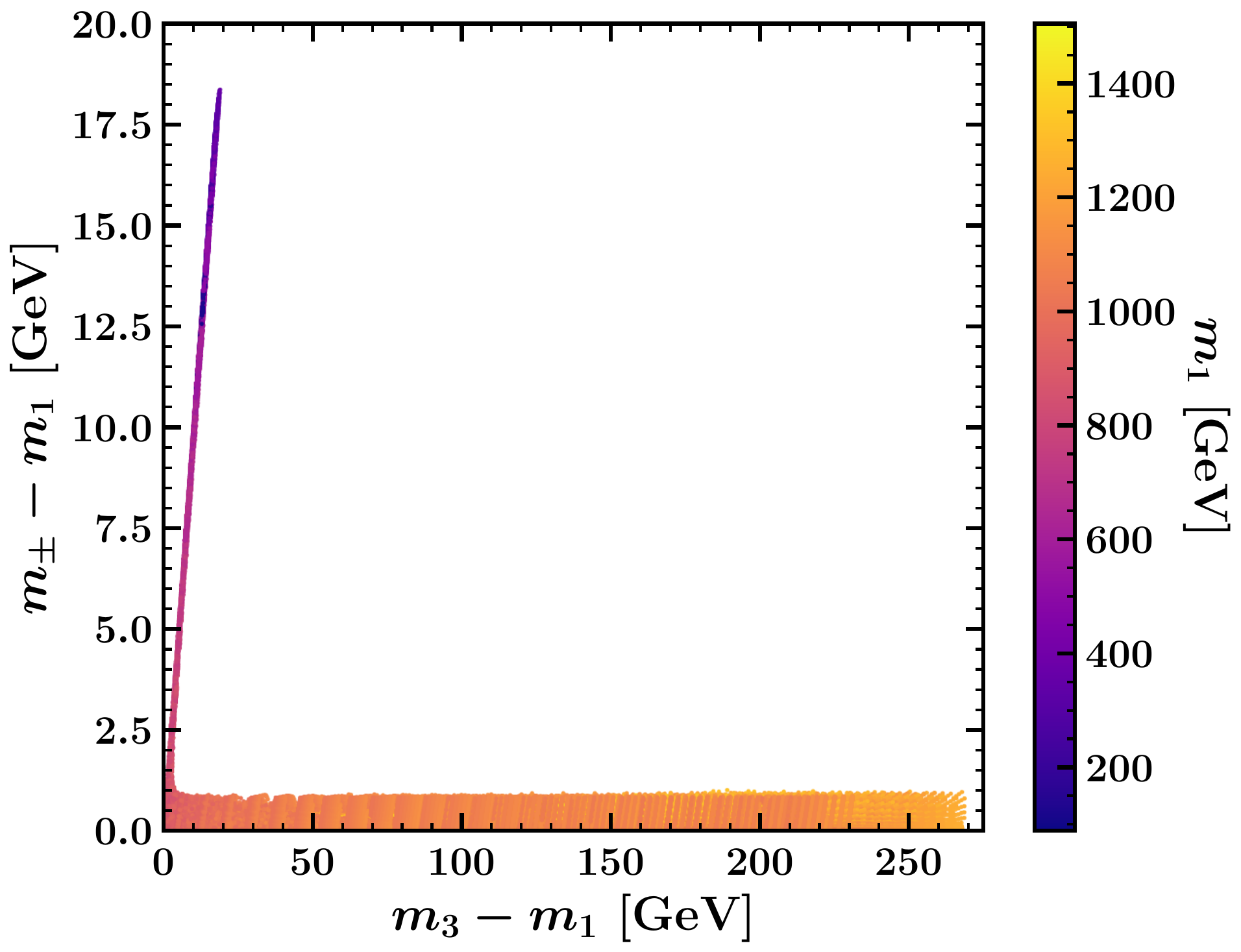}    
\end{subfigure}
\caption{\textbf{Left:} Mass difference between the charged state and the DM $(m_{\pm} - m_{1})$  vs mass difference between second lightest neutral state and the DM $(m_{2} - m_{1})$   \textbf{Right:} Mass difference between the charged state and the DM $(m_{\pm} - m_{1})$ vs mass difference between third lightest neutral state and the DM $(m_3 - m_1)$. In both panels, the color bar indicates the dark matter mass. Here, we show data points that satisfy the relic density condition and have direct detection cross sections beyond the LZ projected sensitivity \cite{LZ:2018qzl}.}
\label{fig:DeltaM_vs_DeltaM}
\end{figure}

\begin{figure}
    \centering
    \includegraphics[width=0.8\textwidth]{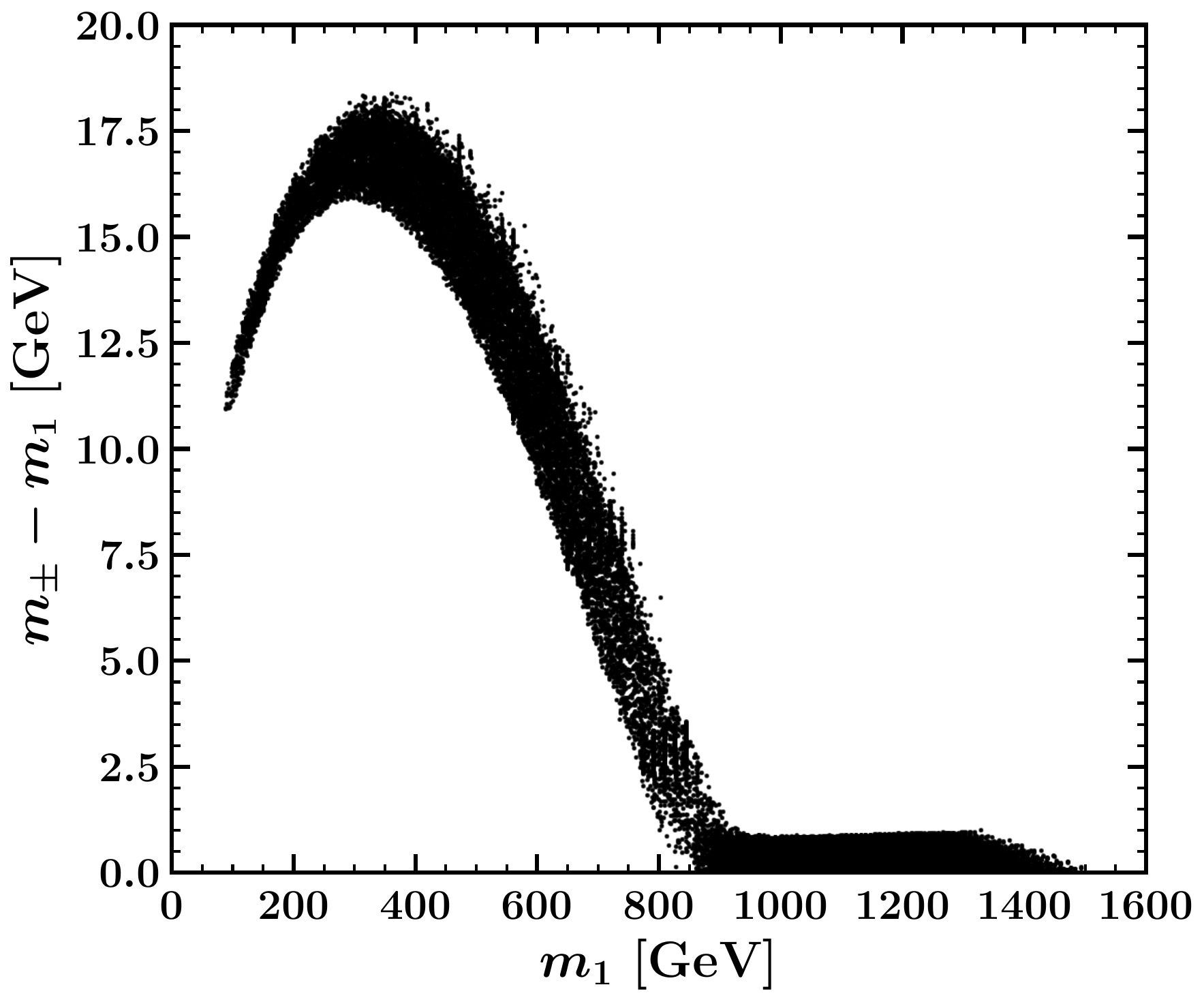}
    \caption{Mass difference between the charged state and the DM as a function of dark matter mass. We show data points that satisfy the relic density and have direct detection cross sections beyond the LZ projected sensitivity \cite{LZ:2018qzl}.}
    \label{fig:MpmMinusM1_vsM1}
\end{figure}

Figure \ref{fig:DeltaM_vs_DeltaM} shows the mass differences between the different states color-coded by the dark matter mass.  Figure \ref{fig:MpmMinusM1_vsM1} shows another view of the mass spectrum specializing to the difference between the mass of the charged state and dark matter that is of particular relevance to collider searches. In both of these Figures, we show data points that satisfy the relic density and have direct detection cross sections beyond the LZ projected sensitivity.  Extending to include points between the current LZ bound and the projected sensitivity does not change the main conclusions.

Each panel in Fig.~\ref{fig:DeltaM_vs_DeltaM} has a region of very near degeneracy between the charged state and the dark matter (the points huddled near the x-axis) as well as a region where this mass splitting extends to several GeV.  To understand what is happening, it is useful to examine the color bar that displays $m_1$  We note that the transition between the two regimes occurs for $m_{1} \sim 850$ GeV. Back in   
Figure \ref{fig:CoupRat_Vs_MassRat_MsCbar} we saw that for points consistent with the relic density, there is a region where $M_S \approx M_D$  that occurs for $m_1 \approx 850$ GeV. So, the entire mass spectrum is very compressed at $m_1 \sim 850$ GeV (when $M_S = M_D$).  Below $m_1 \sim 850$ GeV (when $M_S < M_D$), we have $m_1 \approx M_S$ and $m_3 \approx M_D$.  Numerically, this means that for $M_{S} < M_{D}$, $m_{\pm}$, $m_2$, and $m_3$ are all nearly degenerate and $m_1$ is $\sim 1-20$ GeV lighter.  As mentioned above, the requirement that there are other dark states nearby present in the bath at the time of freeze-out (i.e. coannihilation is effective) means that in this regime, $M_S \gtrsim 0.9M_D$. Above $m_1 \sim 850$ GeV (when $M_S > M_D$), we have $m_1 \approx M_D$ and $m_3 \approx M_S$.  The dark matter with mass $m_1$ and the charged state mass $m_{\pm}$ are degenerate. The other primarily doublet state $m_2$ ranges from nearly degenerate with $m_{\pm}$ to $\sim 1-20$ GeV heavier.  $m_3$ can be much heavier than any of the other states.  The maximum splitting between $m_1$ and $m_3$ in Figure \ref{fig:DeltaM_vs_DeltaM} is limited by our scan region.   
Again, since $\chi_2^0$ and $\chi^{\pm}$ are automatically nearly degenerate with $\chi_1^0$ when $M_D < M_S$, $M_S$ need not be near $M_D$ for effective coannihilation.

Here, we have computed the masses at tree level. Loop level effects are important for the determination of mass splittings between the degenerate states with mass near $M_D$ \cite{Thomas:1998wy}. These splittings play a potentially important role in collider phenomenology (see below), but they are generally irrelevant for cosmological histories.  Temperatures in the early Universe at freeze-out are well in excess of these loop-induced splittings.  

In the following Sections, we explore the implications of this mass hierarchy on the relic density and collider detection prospects of this model.

\subsection{Early universe production}
\label{sec:earlyUniverse}
\begin{figure}
    \centering
    \includegraphics[scale=1.0]{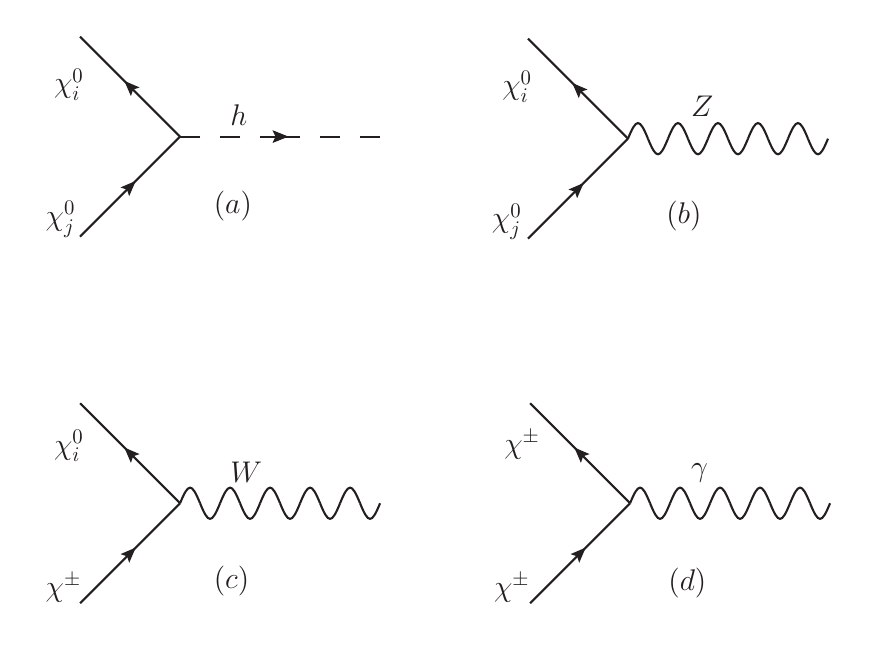}
    \caption{Interaction vertices of the new, dark sector particles with (a) the SM higgs boson, (b) Z boson, (c) W boson, and (d) photon. Relevant vertices for the interaction of the new, dark sector particles with the Standard Model.  Details on the values of these vertices can be found in Eqs.~(\ref{eq:SMcouplings}) and (\ref{eq:SMcouplings2}), and in Appendix \ref{sec:appendix}.}
    \label{fig:Vertices}
\end{figure}

The only interactions between the dark sector and the SM are mediated by the photon and $Z$,  $W$, and Higgs bosons. Figure \ref{fig:Vertices} shows the  vertices for these interactions. These interactions must be sufficient to power a successful freeze-out without inducing too-large direct detection. 

Since the spin-independent and spin-dependent direct detection bounds are strong, vertices (a) and (b) with $i = j = 1$ must be small.  This means that the corresponding annihilation
diagrams with an intermediate  $h$ or $Z$ must also be small. 
Therefore, diagrams involving the $W$ boson or diagrams involving (a) or (b) where both $i$ and $j$ are not equal to $1$ are  important during freeze-out.  All of these diagrams involve at least one non-DM dark sector state.    

It turns out ($\chi^0_1 \chi^0_1 \rightarrow$ anything) is never the dominant annihilation process, and coannhilation dominates. Due to differing amounts of Boltzmann suppression for particles of different mass, the coannhilation rate is sensitive to the mass spectrum like $\sim e^{-x\Delta_{i}}$ where $x \equiv m_1/T$ and $\Delta_{i} \equiv (m_i - m_1)/m_1$ \cite{Griest:1990kh}. This sensitivity to the mass splitting  allows the spectrum predictivity seen in Figure~\ref{fig:MpmMinusM1_vsM1}.

To get a better picture of the processes relevant for freeze-out, it can be helpful to understand the diagonalization of the mass matrix, since the elements of the mixing matrix appear in the couplings between the mass eigenstates and the SM bosons. In Appendix \ref{sec:appendix}, we discuss the diagonalization of the mass matrix and the couplings in detail.  We also explicitly enumerate couplings for a few cases of particular interest. Below, we will discuss the processes of most relevance, and will refer the reader to the Appendix for details as needed. 

Let $DS$ denote a generic dark sector particle and let $f$, $f^{\prime}$ and $V$, $V^{\prime}$ denote SM fermions and gauge bosons respectively.  The interaction $DS\ DS \rightarrow f f^{\prime}$ is the most important freeze-out process for nearly all viable parameter space. The most important class of contributions comes from $\chi^{\pm}\chi^0_i \rightarrow f f^{\prime}$.  Annihilations to gauge bosons, $DS\ DS \rightarrow VV^{\prime}$ also contribute and become comparable to $DS\ DS \rightarrow f f^{\prime}$ in corners of parameter space (for example, in the pure doublet limit we expect annihilation to gauge bosons to dominate, as is well known from the Higgsino limit of the MSSM). Interactions of the form $DS\ DS \rightarrow hV$ also contribute but are subdominant for all parameter space considered.  Note that the DM candidate $\chi_1^0$ itself need not always participate in the freeze-out processes.  Often, freeze-out occurs through coannihilations of heavier dark sector states with other heavier dark sector states (e.g. $\chi^{\pm}\chi^0_2 \rightarrow f f^{\prime}$). The abundance of the dark matter is then ensured by equilibrium between these states and the dark matter.

We have seen above in Section \ref{sec:RelicDensity} that the direct detection and relic density constraints  favor two regimes. The first has small Yukawa couplings.  When $y_i$ are small, direct detection cross sections can be small even for lighter masses and even away from the blind spot(s).  In Appendix \ref{sec:appSmally}, we show the dark sector mass eigenvalues (Eq.~(\ref{eq:SmallY_MassEigenvalues})) and the couplings for the small $y_i$ limit (See Eqs. (\ref{eq:Wcoups_smallY})-(\ref{eq:hcoups_smallY}).)
Even though the direct detection cross sections are tiny, importantly, there are still two $W$ couplings to the $DS$ and one (off-diagonal) $Z$ coupling that 
are of order the electroweak coupling constant $g$.  It is these couplings that are relevant for freeze-out.  Since all of these couplings involve dark sector states beyond the dark matter itself, this bolsters the earlier assertion that coannihilations are important.  Other diagrams that have two dark matter particles in the initial state, with a dark sector particle exchanged in the $t-$channel are subdominant.

\begin{figure}
    \centering
    \includegraphics[width=0.8\textwidth]{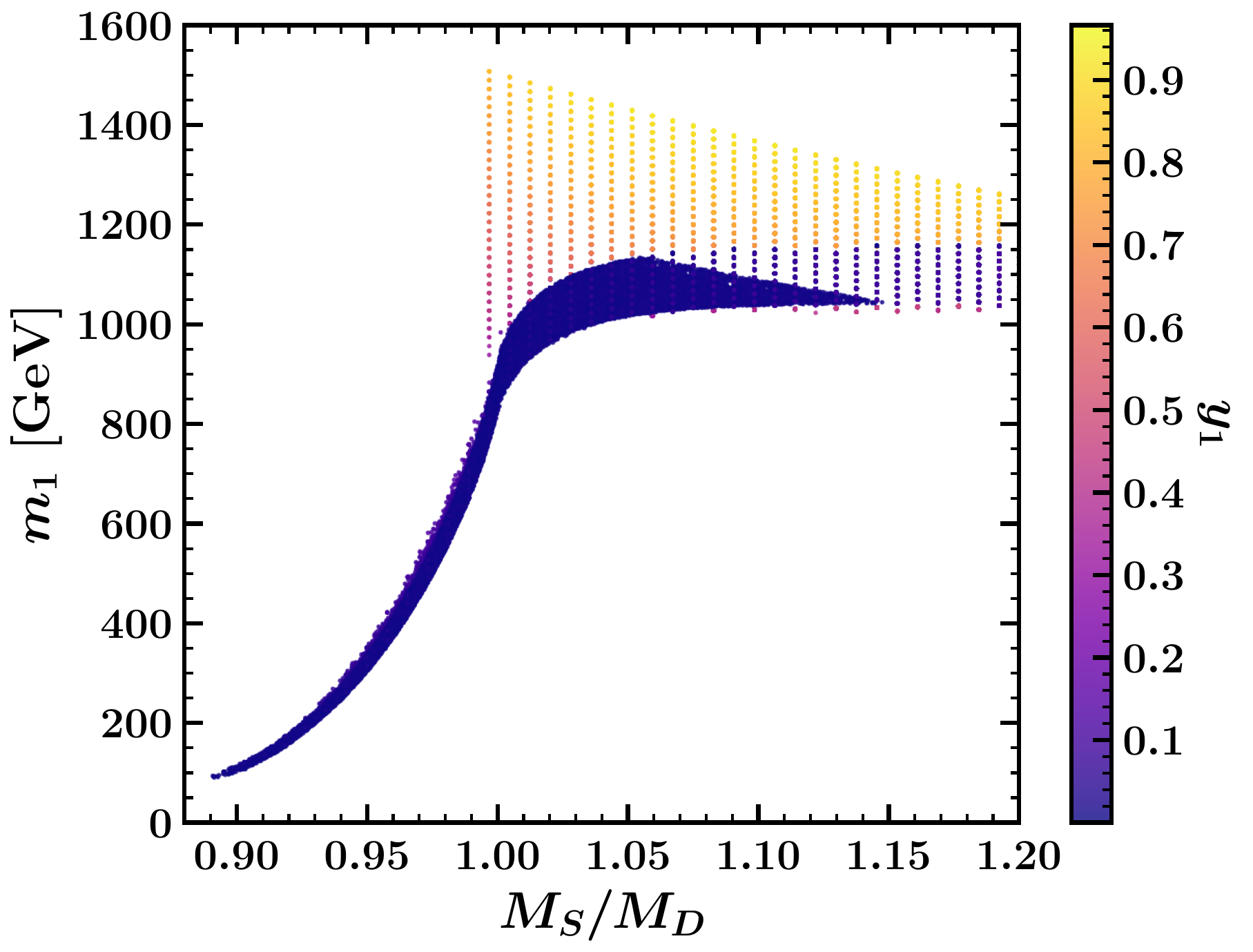}
    \caption{Dark matter mass $m_1$ as a function of the mass parameter ratio $M_S/M_D$, color-coded by the values of $y_1$, for parameter points that yield the correct dark matter abundance and have direct detection cross sections below the LZ projected sensitivity \cite{LZ:2018qzl}.}
    \label{fig:m1vsMSbyMD}
\end{figure}

The second regime is near direct detection blind spot(s).  In this regime,  $y_i$ can be larger because of the associated suppression of the direct detection rates.  Since both the SI and SD direct detection signals must be suppressed,  the double blind spot (with $M_{S} > M_{D}$, $y_2/y_1 = -1$) is of particular interest.  When $y_2/y_1 = -1$, $M_S > M_D$ and $y_1 \sim 1$, the relic density constraint favors $\sim$ TeV scale dark matter.  The cross sections for annihilation to pairs of gauge bosons with mass $M_{V}$ are enhanced by extra factors of $m_1/M_{V}$ when compared to the cross sections to go to pairs of SM fermions, see, e.g., \cite{Nihei:2002ij}.  This explains the relative importance of annihilation to vector bosons in this high-mass regime.  Mass eigenvalues and couplings at the double blind spot are also reported in Appendix \ref{sec:appdoubleBS} (See Eq. \ref{eq:DoubleBS_MassEigenvalues}, and Eqs.~(\ref{eq:Wcoups_doubleBS})-(\ref{eq:hcoups_doubleBS}).)
Once again, although the direct detection cross sections completely vanish, there are still couplings between dark sector particles that are of order the electroweak gauge coupling and mediate freeze-out.  

In fact, for both the small Yukawa and near blind spot regimes, freeze-out largely proceeds these full-strength gauge couplings. So, the remaining viable parameter points smoothly interpolate between these two regimes.  To see how this transition takes place and how close to the double blind spot one needs to sit for different values of the Yukawa couplings see Figure \ref{fig:CoupRat_Vs_MassRat_SmallandLargeY1s}.

Finally, we discuss an implication of allowed large Yukawa couplings close to the double blind spot.  In Fig.~\ref{fig:m1vsMSbyMD}, we have shown the DM mass as a function of $M_{S}/M_{D}$.  For $M_{S}/M_{D} >1$, large  Yukawa couplings are allowed, see Fig.~\ref{fig:CoupRat_Vs_MassRat_SmallandLargeY1s}.  The consequence is that early Universe annihilations can be enhanced.  This, in turn, allows the dark matter to be heavier than in the case ($m_{1} \approx 1100$ GeV) where only gauge interactions are relevant.  For the maximum allowed value of $y_{1}=1$ in our scan, we find dark matter masses as large as 1500 GeV.  The effect of large Yukawa couplings is lessened at larger $M_S/M_{D}$ as mixing is reduced in this regime, see Appendix \ref{sec:appendix}.

Finally, when relying on coannihilation processes to establish the correct relic density, it is necessary to confirm chemical equilibrium is maintained between all dark sector particles.  We have applied a cut to ensure that this equilibrium is indeed maintained.  If it were not, a more careful treatment is required, and coscattering and freeze-in can become important. Discussion of this regime has appeared in \cite{Calibbi:2018fqf}. Analysis of the related singlet-triplet model (which mimics the Wino-Bino system) in this regime appears in Ref. \cite{Alguero:2022inz}. 

\subsection{Singlet-doublet model at the LHC}
\label{sec:colliders}
We next discuss the collider phenomenology of the singlet-doublet model.  For related work see Refs. \cite{Freitas:2015hsa, Calibbi:2015nha, Arganda:2024tqo, ATLAS:2024lda, CMS:2024gyw, ATLAS:2024qxh}. We focus on parameter space beyond the projected reach of LZ and with dark sector states  within reach of the LHC.  The strength of direct detection probes allows sharp statements to be made about both the anticipated production rate for dark sector states at the LHC as well as their decays.

\subsubsection{Couplings and production}
As can be seen above, the remaining viable parameter space potentially within reach of the LHC will have
$0.9 \lesssim M_S/M_D \lesssim 1$, with  singletlike dark matter $\chi^0_1$ with mass $m_1 \simeq M_S$,
and  Yukawa couplings $y_{1}$ and $y_{2}$ with relatively small magnitude.  It can therefore be analyzed using the expressions for the couplings between the dark sector and the SM electroweak bosons in the small Yukawa limit given in the Appendix (Eqs.(\ref{eq:Wcoups_smallY})-(\ref{eq:hcoups_smallY})).

The largest couplings involve
the  doubletlike states $\chi^0_{2,3}$ and $\chi^\pm$, such as $W^\mp-\chi^{\pm}-\chi^0_{2,3}$, $Z/\gamma-\chi^+-\chi^-$, and $Z-\chi^0_2-\chi^0_3$.
All other couplings are small. Therefore, the dominant production mechanisms for this model at the LHC are via intermediate electroweak
bosons which produce some combination of $\chi^{\pm}$, $\chi^0_{2,3}$.

\begin{figure}
    \centering
    \includegraphics[scale=0.5]{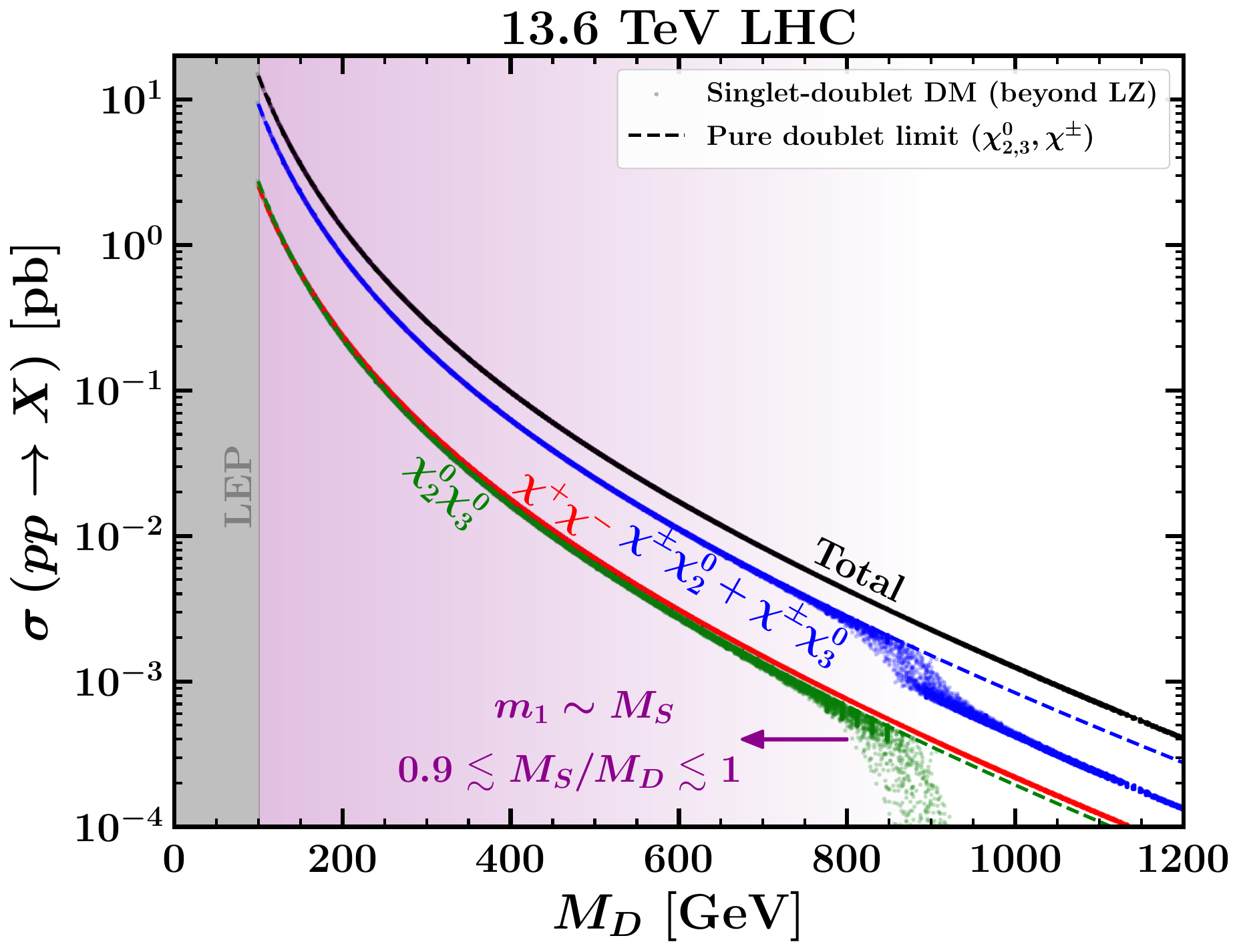}
    \caption{Dark sector pair production cross sections at the LHC with $\sqrt{s} = 13.6$ TeV  as a function of the doublet mass parameter $M_D$.  Points shown satisfy the observed relic density and have direct detection cross sections beyond  projected LZ sensitivity.
    Dashed lines show  corresponding cross sections in the pure doublet limit.
    The purple-shaded region (faded to the right) highlights points potentially within reach of the LHC.   They satisfy $0.9 \lesssim M_S/M_D \lesssim 1$ with the dark matter mass $m_1 \simeq M_S$, see Fig.~\ref{fig:MassSpectSchematic}.
    Different colors denote different (combined) final states: $\chi^+\chi^0_2, \chi^-\chi^0_2, \chi^+\chi^0_3, \chi^-\chi^0_3$ (blue), $\chi^+ \chi^-$ (red), $\chi^0_2 \chi^0_3$ (green), and all final states (black) including the ones that do not occur in the pure doublet case, i.e., $\chi^+\chi^0_1, \chi^-\chi^0_1$ and $\chi_i^0\chi_j^0$ with $i \leq j$ and $i, j \ne 2, 3$.
    The vertical gray shaded region is the 95\% CL exclusion from the Large Electron-Positron (LEP) collider \cite{L3:2001xsz}.}
    \label{fig:sigmaLHC}
\end{figure}

Furthermore, pair production cross sections for the dominant final states are expected to be very close to those in the pure doublet limit, where $\chi^0_{2,3} \sim \frac{\overline{N} \pm N}{\sqrt{2}}$ (see text surrounding Eq.~(\ref{eq:rotatedbasis})) are nearly mass-degenerate at $M_D$ along with $\chi^\pm$.
This is illustrated in Figure~\ref{fig:sigmaLHC}, where the scatter points correspond to the points that reproduce the correct dark matter abundance and lie beyond the projected LZ reach, and the dashed lines correspond to the cross sections in the pure doublet limit.
The purple-shaded region faded to the right of the plot highlights points potentially within reach of the LHC.
The different colored curves denote different (combined) final states: $\chi^+\chi^0_2, \chi^-\chi^0_2, \chi^+\chi^0_3, \chi^-\chi^0_3$ (blue), $\chi^+ \chi^-$ (red), $\chi^0_2 \chi^0_3$ (green), and all final states (black) including the ones that do not occur in the pure doublet case, i.e., $\chi^+\chi^0_1, \chi^-\chi^0_1$ and $\chi_i^0\chi_j^0$ with $i \leq j$ and $i, j \ne 2, 3$.  Note that for the region potentially accessible at the LHC, collider production is well described by the pure doublet limit. The vertical gray shaded region ($M_D \lesssim 101.2$ GeV) corresponds to the 95\% CL exclusion from the Large Electron-Positron (LEP) collider \cite{L3:2001xsz}, inferred from the nonobservation of heavy charged leptons.

For viable parameter space likely near the edge of the kinematic reach of the LHC, the green and blue points can begin to deviate from the corresponding pure doublet cross sections.  This occurs as the dark matter candidate $\chi^0_1$ acquires a larger doublet component.  Total cross sections match the pure doublet limit when summing over all final states (black points), or even just over the additional final states arising through singlet-doublet mixing within each color-coded final state category (i.e., blue points plus $\chi^+ \chi^0_1, \chi^- \chi^0_1$, and green points plus $\chi^0_i\chi^0_j$ with $i\leq j$ and $i,j \ne 2,3$).
This is because the primarily singlet state inherits gauge interactions entirely through mixing with the doublet.

\subsubsection{Decays}

To understand the potential collider signatures of the model, we now discuss the decays of those particles produced with significant cross section ($\chi^{\pm}$, $\chi^0_2$ or $\chi^0_3$).  We seek to understand whether these particles are long-lived and whether there can be cascade decays or whether dark sector states immediately decay to $\chi_{1}^0$. 

For parameter space within LHC reach, all mass differences in the dark sector are less than $M_W$, so three-body decays are relevant.  The dominant three-body decays are mediated by off-shell $W$ and $Z$ bosons; off-shell Higgs-mediated decays are suppressed by small SM Yukawa couplings.

Three-body decay rates scale as $(\Delta m)^5$ with $\Delta m$ the mass difference between the initial and final dark sector states.  As seen in Figure \ref{fig:MassSpectSchematic}, for $M_{S} < M_{D}$, there are three nearly degenerate (mostly doublet) dark sector states with masses $m_2, m_{\pm}, m_3\approx M_D$ and a somewhat lighter DM with mass $m_1\approx M_S$. The mass difference between the DM $\chi_1^0$ and the next lightest state ($\Delta m$) is between $\sim5-20$ GeV (see Figures~\ref{fig:DeltaM_vs_DeltaM} and \ref{fig:MpmMinusM1_vsM1}) and is typically much larger than the mass differences between the nearly degenerate states clustered near $M_D$.  These splittings are typically of ${\mathcal O}(100 \; \rm{MeV})$ or less. The strong $(\Delta m)^5$ scaling favors decays directly to the dark matter (as opposed to a cascade).  We confirmed these decays are typically dominant by using appropriately modified expressions for the three-body decay expressions of MSSM neutralinos from Ref. \cite{Djouadi:2001fa}. 

Also, for most of our parameter space, the relatively large values of $\Delta m \sim 5-20$ GeV mean that the decays of the heavier states are prompt.  However, for sufficiently small values of the Yukawa couplings, it is possible for the decays of the heavy neutral states to result in displaced vertices.  This  might lead to displaced lepton pairs.  Typically, for the parameter space studied in this work, the lifetimes will be less than a second, and therefore not interfere with BBN. 

For the charged state, the decays are always prompt.  This is due to the existence of decays involving the pion within the doublet sector, i.e. $\chi^0_3 \rightarrow \chi^{\pm} \pi^{\mp}$ or $\chi^{\pm} \rightarrow \chi_2^0 \pi^{\pm}$ \cite{Thomas:1998wy}.  Instead of $(\Delta m)^5$, these decays scale as $f_{\pi}^2(\Delta m)^3$. Again, these decays are typically subdominant to direct decay to $\chi^0_1$ for parameters of interest.  However, in the limit of very small Yukawa couplings these decays become important.  As in the case of the pure Higgsino of the MSSM, even the radiative splitting between the charged and the $\chi_{2}^0$ is sufficiently large to admit these prompt decays to charged pions.  For small values of the Yukawa couplings, mixing with the singlet does not typically change that fact.

In specific corners of parameter space, it is possible to overcome the $(\Delta m)^5$ scaling.  For example, when $y_2 = -y_1$ the $\chi^0_3-\chi^0_{1}-Z$ coupling vanishes\footnote{We find that the loop induced decay $\chi^0_3 \rightarrow \chi^0_1\gamma$ also vanishes for this coupling combination. \cite{Haber:1988px, Baum:2023inl}}.  In this case,  $\chi^0_{3}$ could decay first to a $\chi^0_{2}+f\overline{f}$ (or possibly $\chi^0_{2} \pi^{0}$ or $\chi^0_{2} \gamma$ depending on the $\chi^0_{2}-\chi^0_{3}$ mass splitting). Due to the small mass splitting, the particles emitted in the $\chi^0_3 \rightarrow \chi^0_2$ decay tend to be quite soft.  The $\chi^0_{2}$ can subsequently decay to the dark matter.  In the limit where the $\chi^0_3 \rightarrow \chi^0_2$ decay is very slow in addition to the $\chi^0_3-\chi^0_{1}-Z$ coupling being turned off, late time cosmological constraints from either BBN or the cosmic microwave background could be relevant.

Taken together with the production cross sections of the previous section, this discussion of decays means the collider phenomenology is well specified.  The final states shown as blue points in Fig.~\ref{fig:sigmaLHC}  decay via an off-shell $Z$ and an off-shell $W$.  This can yield contributions to a state with three fairly soft leptons. The mass splitting relevant for the decays can  be seen in  Fig.~\ref{fig:MpmMinusM1_vsM1}, recalling the states produced $\chi_{2,3}^{0}, \chi^{\pm}$ are all nearly degenerate at $M_{D}$; see also Fig.~\ref{fig:DeltaM_vs_DeltaM}.  The final state shown as green in Fig.~\ref{fig:sigmaLHC}  decays via a pair of off-shell $Z$ bosons, which could in principle yield four-lepton final states, but because of the relatively small cross section and small $Z$  branching ratio to leptons, this channel suffers from small statistics.

\section{Renormalization Group Focusing}
\label{sec:focusing}

As argued in previous sections, to achieve sufficient annihilation in the early universe while evading direct detection constraints requires special relationships between the model parameters. It is possible that these relationships are accidental, but an interesting alternative is that special relationships might hold in the infrared (IR) because of a renormalization group focusing effect \cite{Kearney:2013xwa}. That is, a relatively wide range of initial conditions in the ultraviolet (UV) might result in special relationships in the IR. 

In the present context, we are interested in whether renormalization group evolution can focus us into regions of parameter space near the blind spots and/or where coannihilation is possible. That is, our interest is in whether RG focusing causes parameters to evolve into regions with large early-universe annihilation cross sections but suppressed direct detection cross sections.  In this Section we study whether RG focusing can reproduce these qualitative features; we do not necessarily seek to reproduce specific points in parameter space that precisely satisfy the relic density constraint. We find this  focusing is often difficult to achieve, but there are regions of parameter space where it could be of relevance.

Our analytic analysis below is carried out at the one-loop level.  This is sufficient for a qualitative picture.  In our numerical results, we incorporate two-loop effects.

We begin by defining the following ratios
\begin{equation}
     r_m \equiv \frac{M_S}{M_D} \qquad \qquad r_y \equiv \frac{y_2}{y_1}.
\end{equation}
In terms of these ratios, the blind spots (Eqs. (\ref{eq:HBS})and (\ref{eq:ZBS})) may be written 
\begin{equation}
    r_y^\text{SD-BS} = \pm 1  \qquad (\text{Z blind spot})
\end{equation}
for the $Z$ (i.e. spin-dependent) blind spot and 
\begin{equation}
    r_y^\text{SI-BS} = -r_m\left(1\pm\sqrt{1-r_m^2}\right)^{-1} \qquad (\text{Higgs blind spot}; \; r_m < 1)
\end{equation}
for the Higgs boson (i.e. spin-independent) blind spot when $r_m < 1$.  When $r_m > 1$, $r_y = -1$ becomes both a spin-independent and spin-dependent blind spot.

\begin{equation}
    r_y^\text{DBS} = -1 \qquad (\text{Double blind spot}; \; r_m > 1)
\end{equation}

We use \textsc{SARAH} to calculate the one-loop beta function for each of these ratios.  The result for $r_y$ is 
\begin{equation}\label{eq:betary}
    \beta_1 (r_y) = \frac{3A}{2} r_y(y_1^2 - y_2^2) = \frac{3A}{2}r_y y_1^2(1-r_y^2)
\end{equation}
where $A \equiv 1/16\pi^2$ is the loop factor.  We note that when $|y_1|$ and $|y_2|$ are similar  (or equivalently $|r_y| \sim 1$) then $r_y$ evolves slowly.  The stationary points of this beta function are
\begin{equation}\label{eq:rystar}
    r^*_{y} = 0, \pm 1.
\end{equation}
While it may be possible to have focusing to $r^{\ast}_{y}=0$, it is not of particular interest because it does not live near a blind spot.  
On the other hand, the presence of the stationary points at $\pm 1$ reflects an enhanced custodial symmetry.  Once the model departs from this point of enhanced symmetry, it would be surprising if RG flow restored this symmetry.  We verify this intuition by exploring perturbations around these points.  We write $r_y = r_y^* (1 + \delta_y)$ 
and then find the evolution for $\delta_y$.  If the value of $\delta_y$ were to go to zero as energy decreases, then it is possible to focus to the these stationary points in the IR.

We have

\begin{equation}
    \beta (\delta_y) = -\frac{3 A}{2} y_1^2 \delta_{y} \left(1 + \delta_{y}\right) \left(2 + \delta_{y}\right).
    \qquad \left(r_y^* = \pm 1; \ r_y = r_y^* (1 + \delta_y) \right)
\end{equation}
For small $\delta_{y}$, $\beta(\delta_{y})$ and $\delta_{y}$ have the opposite sign.
So, the absolute value of $\delta$ increases with decreasing energy scale, and RG evolution moves the couplings away from the point of enhanced symmetry in the IR.  

Now, we apply a similar analysis to the ratio $r_m$. The one-loop beta function for $r_m$ is given by

\begin{equation}\label{eq:betarm}
    \beta_1 (r_m) = A \left(2 r_y y_1^2\left(4-r_m^2\right) + B r_m\right)
\end{equation}
where, as above, $A \equiv 1/16\pi^2$ is the loop factor,
and we define $B \equiv \frac{3}{2} y_1^2\left(1 + r_y^2\right) + \frac{9}{10} \left(g_1^2 + 5 g_2^2\right)$.
Since $S$ and $D$ have different gauge quantum numbers, $M_S$ and $M_D$ evolve differently under the renormalization group, even in the case when $y_1$ and $y_2$ are set to $0$.  This effect is encompassed by the last term of Eq.~(\ref{eq:betarm}) above. Setting $\beta_1 (r_m) = 0$ we find pseudostationary points

\begin{equation}\label{eq:rmstar}
    r_m^* = \frac{B}{4 y_1^2 r_y} \mp \sqrt{4 + \left(\frac{B}{4 y_1^2 r_y}\right)^2}
\end{equation}
where, as above, $B \equiv \frac{3}{2} y_1^2\left(1 + r_y^2\right) + \frac{9}{10} \left(g_1^2 + 5 g_2^2\right)$. We work in a convention where $r_m > 0$.  So, for $r_y < 0$ we take the minus sign in front of the square root and for $r_y > 0$ we take the plus sign in front of the square root. (If $r_y = 0$, the only solution is $r_m = 0$). Note $r_m^*$ is not a true fixed point.  It is a function of couplings that themselves experience RG evolution.  In general, the  evolution of the mass ratio in Eq.~(\ref{eq:betarm}) will not match the evolution of the zero of the beta function, i.e. $r_m(\mu) \neq r_m^*(\mu)$.  In other words, if $r_{m} = r_m^*$ at a some energy scale, this relation will not precisely hold as evolution continues into the IR. However, evolution away from this point will be slow, as it will be formally of higher loop order.  
What about evolution towards this pseudostationary point in the first place? For a particular scale $\hat{\mu}$, $r_m (\hat{\mu}) > r_m^*(\hat{\mu}) \implies \beta(r_m(\hat{\mu})) > 0$ and $r_m (\hat{\mu}) < r_m^*(\hat{\mu}) \implies \beta(r_m(\hat{\mu})) < 0$.  This means that, provided the evolution of $r_m^*(\mu)$ is not too rapid, $r_m$ will move toward $r_m^*$ in the IR.

Below,  we explore applications of the above RG analysis to some particular cases of interest. For each case, we explore the range of parameter space in the UV where, after undergoing RG evolution, we end up with features of interest (i.e. small direct detection, and large early universe annihilation) at the electroweak scale.

\subsubsection{Small Yukawa couplings}

As we have seen, small $y_1$ and $y_2$ suppress the direct detection cross sections.  In this case, it is possible to stray further from the blind spots while maintaining consistency with current experimental bounds.  In this regime RG focusing plays no role.  When the Yukawa couplings are small $\beta_1 (r_y) \sim \mathcal{O}(y_1^2)$ and $\beta_1(r_m) \sim 0.01 r_m + \mathcal{O}(y_1^2)$.  The factor of 0.01 comes from plugging in the values of the SM gauge couplings.  Moreover, when the Yukawa couplings are small, the running for $y_1$ and $y_2$  themselves are suppressed.  Since RG evolution is slow, this means that if one starts out in the UV in a particular region of parameter space (e.g. with small Yukawa couplings or near a blind spot) these relations will be approximately preserved under RG evolution.  However, if one starts away from the special regions of parameter space, there will not be any significant focusing into the special regions in the IR.

\subsubsection{Double blind spot}

The area of parameter space where $r_y = -1$ and $r_{m} >1$ is of particular interest because both the spin-independent and spin-dependent direct detection cross sections are suppressed.  As mentioned above, $r_y = -1$ is a point of enhanced symmetry of the model and is a fixed point under RG evolution.  We can therefore consider the case where the theory respects this symmetry (i.e. fix $r_{y}=-1$) and then inquire about the behavior of $r_{m}$.

\begin{figure}
    \centering
    \includegraphics[scale=0.5]{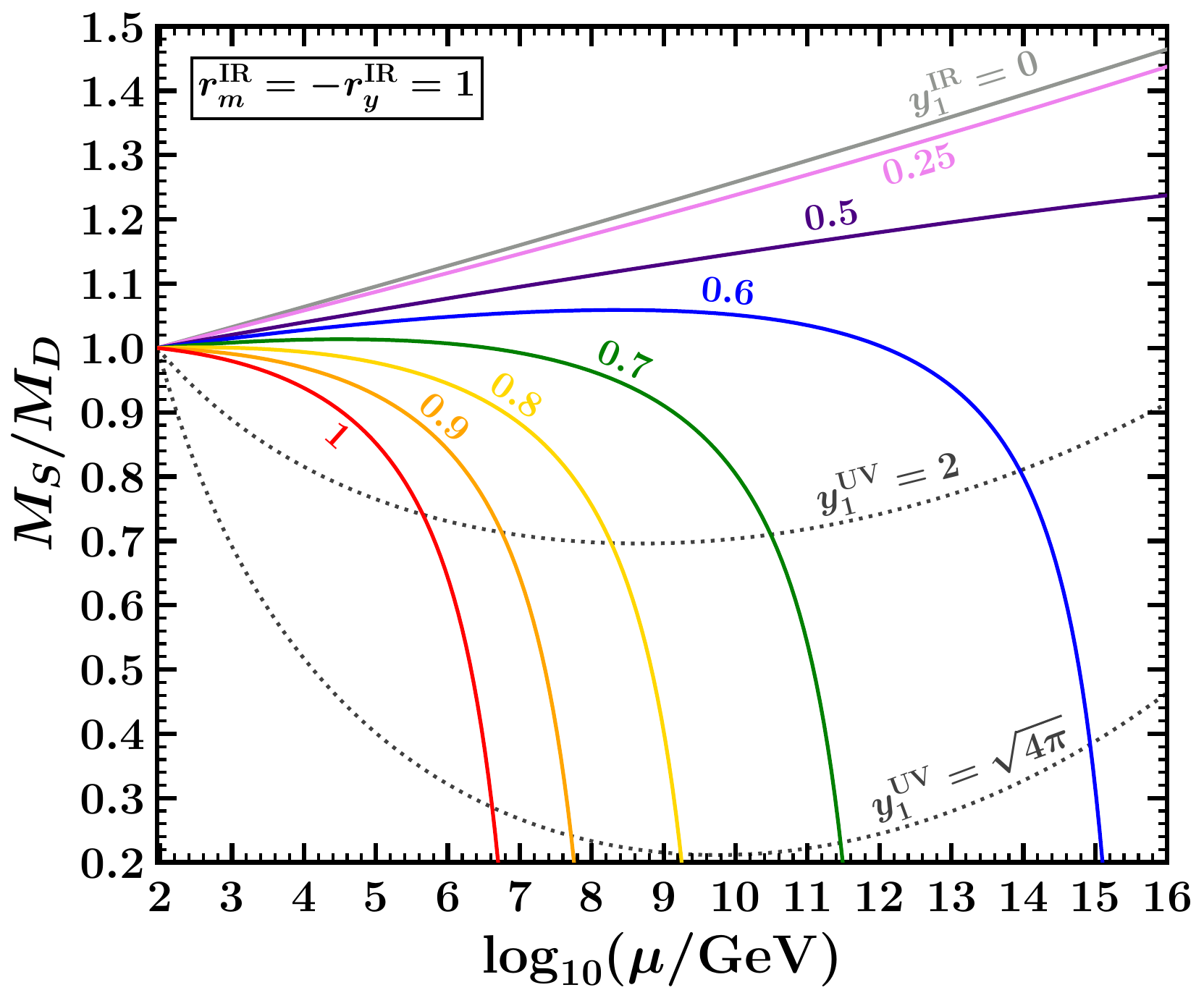}
    \caption{RG evolution of the mass ratio starting from $r_y^\text{IR} = -1$ and $r_m^\text{IR} = 1$. The different colors correspond to different values of $y_1^\text{IR}$.
    The dotted lines indicate contours along which the RG-evolved $y_1$ reaches $2$ and $\sqrt{4\pi}$ in the UV, as labeled.}
    \label{fig:RGFocusing_DoubleBS}
\end{figure}

Although $r_m$ does not have a true stationary point, we expect there should be some focusing due to the existence of $r_m^*$ - see Eq. (\ref{eq:rmstar}).  In Figure~\ref{fig:RGFocusing_DoubleBS}, we fix $r_y^\text{IR} = -1$ and $r_m^\text{IR} = 1$ and then RG evolve the mass ratio to higher energies.  By examining how values well away from $M_{S}/M_{D}=1$ in the UV  end up with $M_{D}/M_{S}\simeq1$ in the IR, we can gain perspective on the amount of focusing that occurs.  The different colored lines represent different values of $y_1^\text{IR}$.  Since we are fixed at the double blind spot in the IR, a large range of $y_1^\text{IR}$ values are compatible with experimental constraints (see Figure \ref{fig:CoupRat_Vs_MassRat_SmallandLargeY1s}).  

When starting from a large $y_1$ value in the IR, it is possible to run into a Landau pole before reaching the Planck scale. For this reason, we show the contours (dotted lines) along which the RG-evolved $y_1$ reaches $2$ and $\sqrt{4\pi}$ in the UV, with the intersection of the solid and dotted lines indicating the cutoff scale at which the coupling becomes nonperturbative.
Above the cutoff scale, we would require a UV completion that would be responsible for setting the initial conditions of the running.\footnote{For the largest values of the Yukawa couplings in the IR, the Landau pole appears before the grand unification scale. Then, improved gauge coupling unification properties cannot be used as a motivation for the singlet-doublet model. There is tension between this focusing mechanism and  unification.}

Interestingly, in Figure~\ref{fig:RGFocusing_DoubleBS} we see that for $y_1^\text{IR}\gtrsim 0.6$ it is possible to start out with a relatively small ratio of $M_S$ and $M_D$ in the UV but end up with $M_S \gtrsim M_D$ in the IR.  This is significant: the RG evolution focuses the values of the model into a region where coannihilation is possible.

As a side note,  Figure~\ref{fig:RGFocusing_DoubleBS} also indicates the possibility of starting out with a larger ratio of $M_S$ and $M_D$ in the UV and evolving toward $r_m= 1$ in the IR.  This focusing is less important from a coannihilation perspective because we always have coannihilation when $M_D < M_S$.  However, it does provide some extra motivation to study the region where $1 \leq r_m \leq 1.2$ that we have analyzed in this work.  That said, it is perfectly possible to start with values of $M_{D} < M_{S}$ in the UV, and also end with this relationship in the IR without resorting to focusing.  The result is a (generalized) version of nearly pure Higgsino dark matter with mass near a TeV.

\subsubsection{Spin-independent blind spot with $r_m < 1$}

\begin{figure}
    \centering
    \includegraphics[scale=0.5]{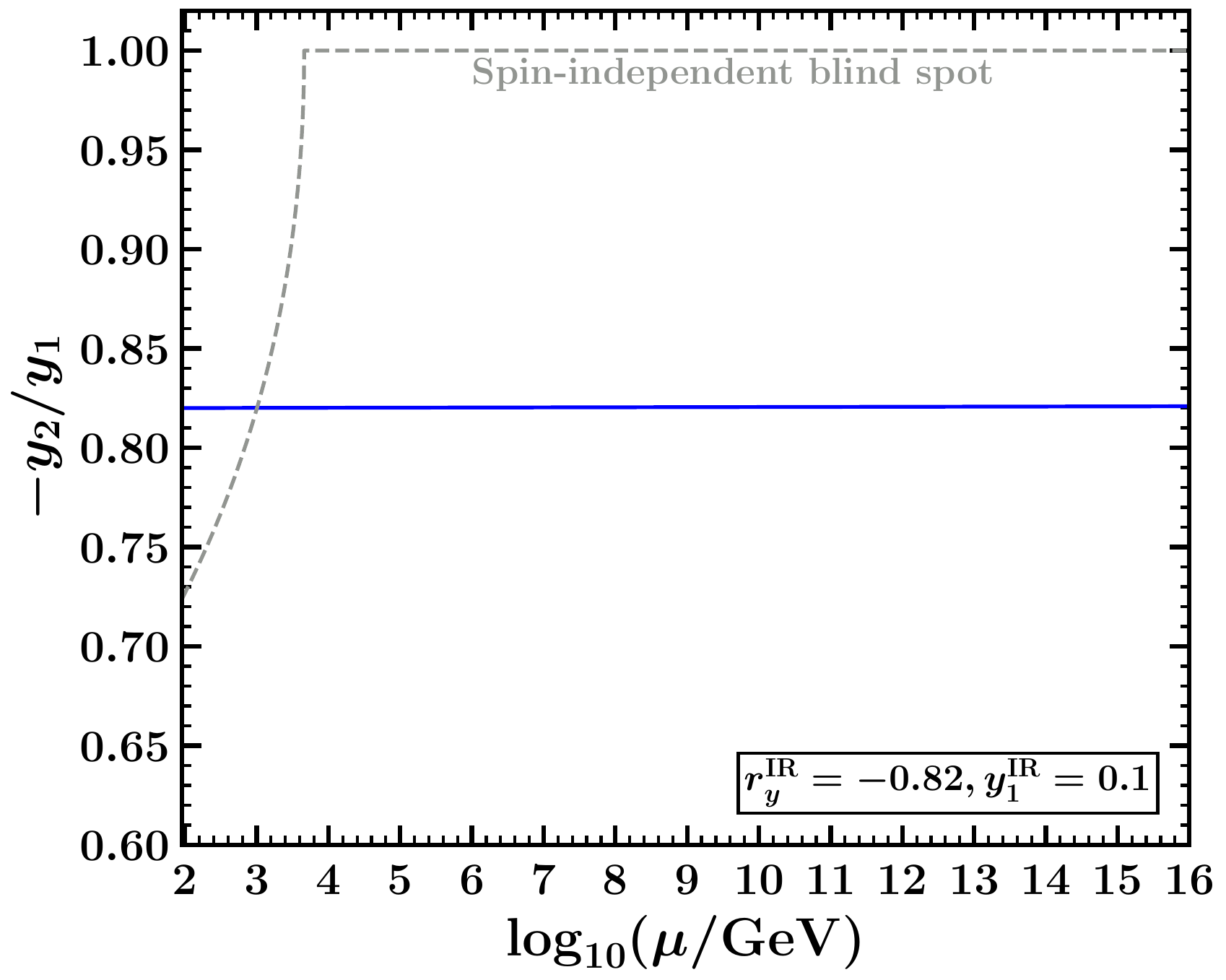}
    \caption{Evolution of the coupling ratio required to sit at the spin-independent blind spot $r_y^\text{SI-BS} (r_m(\mu))$ as compared to the RG evolution of the coupling ratio $r_y (\mu)$.  These two quantities could end up near each other at low energies but it is not guaranteed.}
    \label{fig:RGFocusing_SI_RM}
\end{figure}

As discussed in Section \ref{sec:bs}, when $r_m < 1$, the spin-independent and spin-dependent blind spots no longer coincide.  Instead, the value of $r_y$ corresponding to the spin-independent blind spot becomes a function of $r_m$ (see Eq.~(\ref{eq:HBS})). Since the SI direct detection experimental bounds are especially stringent, evading them prioritizes parameters near the SI blind spot.  But since $r_m$ experiences RG evolution as a function of energy scale $\mu$, the blind spot condition also evolves as a function of $\mu$.  The RG evolution of the blind spot  (i.e. $r_y^\text{SI-BS} (r_m(\mu))$) and of the ratio of couplings $r_y (\mu)$ do not coincide. So, couplings that satisfy the blind spot relation at high energy will not maintain this relation down to low energies.  Moreover, since the spin-independent blind spot when $r_m < 1$ does not necessarily coincide with a stable fixed point of $r_y$, focusing to this point under RG evolution should not be expected.  Ending up near the spin-independent blind spot in the IR would be a coincidence.  Figure \ref{fig:RGFocusing_SI_RM} shows an example of how this might occur. It shows $r_y^\text{SI-BS} (r_m(\mu))$ and $r_y (\mu)$ evolution for the particular choice $r_m^\text{IR} = 0.95$, $M_D = 300$ GeV, $y_1^\text{IR} = 0.1$ and $y_2^\text{IR} = -0.082$.  These choices for $r_m$, $M_D$, and $y_1$ are chosen to be in parameter space of interest as shown in Section \ref{sec:results}.  
 
The spin-independent blind spot in Figure \ref{fig:RGFocusing_SI_RM} is frozen at  $r_y = -1$ for many decades in mass scale, only beginning to evolve around $\sim$ TeV scale. Until this scale, $r_m >1$, and so the blind spot is fixed at $-y_{2}/y_{1}=1$. It is only once $r_m<1$ that the blind spot evolution kicks in, and then $r_y^\text{SI-BS} (r_m(\mu))$ goes like some negative order one number times the mass ratio beta function, $-\mathcal{O}(1)\times \beta(r_m)$. 
Meanwhile, the ratio of Yukawa couplings evolves very slowly.  This is because the Yukawa couplings in this area of parameter space are $\mathcal{O}(0.1)$, and the $\beta$ function for $r_y$ in Eq.~(\ref{eq:betary}) is proportional to $y_1^2$.
As seen in Figure \ref{fig:RGFocusing_SI_RM} and for most of parameter space $|\beta(r_m)| \gg  |\beta(r_y)|$.  So, over the last few decades of evolution, the blind spot steeply descends from $-r_y = 1$ while $r_y(\mu)$ continues to evolve slowly. It is possible (but would be coincidental) for these to nearly coincide near the TeV scale as shown.  This emphasizes that this blind spot requires a tuning of the underlying parameters.

\subsubsection{RG focusing summary}
In this section, we have studied the one-loop RG evolution of the mass ratio $r_m$ and the coupling ratio $r_y$. 
In the case of $r_y$, there is a fixed point and we studied its stability.  In the case of $r_m$, we argued that, although it does not have a true fixed point, it is possible to get RG focusing.  

We applied our analysis to three areas of parameter space of interest -  (1) the small Yukawa coupling limit, (2) the double blind spot, and (3) the spin-independent blind spot when $r_m < 1$.  For (3), RG focusing does not enforce the coupling ratios required to end up in the spin-independent blind spot when $r_m < 1$, and ending up at the blind spot is a coincidence.  In case (1), Yukawa couplings are small, and no significant focusing occurs.  Case (2) is of the most interest. When $r_y = -1$ the symmetry of the model is enhanced and the coupling ratio exhibits a stationary point.  This means that if couplings obey $r_y = -1$ in the UV, this ratio is maintained under RG evolution.  We showed that in this case it is possible for $M_{S}$ to take on values well below $M_{D}$ in the UV but end up larger than $M_{D}$ in the IR. Then, the model could inhabit the double blind spot in the IR.

\section{Conclusion}

We have discussed the remaining parameter space for the WIMP singlet-doublet dark matter model. 
To evade the increasingly stringent direct detection bounds, the model is forced to either have small Yukawa couplings or live near a blind spot.  In these regimes of parameter space, coannihilation (and thus a compressed mass spectrum) is required to achieve the correct relic density.  The direct detection cross section can lie beyond the projected sensitivity of LZ and even below the neutrino fog.  We show how these hard-to-detect areas of parameter space will appear at the LHC. 

The viability of the singlet-doublet model relies on special relationships between the model parameters.  In the last part of this work we studied the RG evolution of the model parameters and showed that, for large values of the Yukawa couplings it is possible to focus the mass ratio $M_{S}/M_{D} \rightarrow 1$ in the IR and end up near the double blind spot.  This could be of particular interest if there were a reason to favor values of $M_{S} < M_{D}$ in the UV.  Another perfectly viable scenario is for $M_{D} < M_{S}$ in both the UV and IR, where the model largely mimics the pure Higgsino of the MSSM.  This scenario is difficult to probe at both the LHC and direct detection experiments but might be in reach of a high-energy lepton collider.

This publication made use of JaxoDraw \cite{Binosi:2003yf}.  This work made use of \textsc{webplotdigitizer}\footnote{\url{https://automeris.io/WebPlotDigitizer/}}, \textsc{numpy} \cite{Harris:2020xlr}, \textsc{scipy} \cite{Virtanen:2019joe}, and \textsc{matplotlib} \cite{Hunter:2007ouj}. 

\begin{acknowledgments}
We thank Stephen P. Martin for helpful comments.
This research was supported in part through computational resources and services provided
by Advanced Research Computing (ARC), a division of Information and Technology Services
(ITS) at the University of Michigan, Ann Arbor.
This work is supported in part by the Department of Energy under Grant No. DE-SC0007859.

\end{acknowledgments}

\section*{Data Availability}
The data that support the findings of this article are openly available \cite{websiteFordata}.

\appendix

\section{Mass Matrix Diagonalization and Couplings} \label{sec:appendix}

The neutral sector of the singlet-doublet model consists of three states.  The mass matrix for these states is given by
\begin{equation}
    \mathcal{M}_N = \begin{pmatrix}
        M_S & \frac{y_1 v}{\sqrt{2}} & \frac{y_2 v}{\sqrt{2}}e^{i\theta}\\
        \frac{y_1 v}{\sqrt{2}} & 0 & M_D \\
        \frac{y_2 v}{\sqrt{2}}e^{i\theta} & M_D & 0
    \end{pmatrix}
    ,
\end{equation}
where, as in the main text, $M_S$, $M_D$, $y_1$ and $y_2$ are our model parameters, $v$ is the standard model vacuum expectation value, and $\theta$ is a phase.   In this work, we specialize to the case $\theta = 0$, and we choose a convention such that $M_S, M_D, y_1 \in \mathbb{R}^{+}$ and $y_2 \in \mathbb{R}$. To determine the mass eigenstates in the neutral sector and the couplings of these states to the SM Higgs and gauge bosons, we need to diagonalize this matrix. 

The mass eigenvalues $\lambda$ satisfy
the characteristic equation $|\mathcal{M}_N - \lambda I| = 0$, which yields
\begin{equation} \label{eq:characteristiceq}
    (\lambda - M_S)(\lambda - M_D)(\lambda + M_D) = (\lambda - M_D)\frac{(y_1 - y_2)^2 v^2}{4} + (\lambda + M_D)\frac{(y_1 + y_2)^2v^2}{4}.
\end{equation}
Generically, solutions to this equation are complicated cubic roots.
However, we focus on a few key limits motivated by finding the parameter space that avoids the very strong direct detection bounds.
These limits include the small Yukawa coupling regime and regions near the blind spots, where the dark matter couplings to the $h$ and/or $Z$ boson vanish.
In these limits, the expressions for the masses and couplings greatly simplify.
A discussion of parameter space with small Yukawa couplings appears in \ref{sec:appSmally}.

First, we consider the blind spots.
The blind spots in Eqs.~(\ref{eq:ZBS})-(\ref{eq:DoubleBS}) correspond to the following special cases
of the characteristic equation:
\begin{alignat*}{3}
    &\text{i) } && \lambda = M_S: \quad 
    && \frac{y_2}{y_1} = -\frac{M_S}{M_D} \left( 1 \pm \sqrt{1 - \left( \frac{M_S}{M_D} \right)^2} \right)^{-1}
    ,
    \quad \text{($h$ blind spot; $M_S < M_D$)} \\
    &\text{ii) } && \lambda = M_D: \quad 
    && \frac{y_2}{y_1} = -1
    ,
    \quad \text{($Z$ blind spot or double blind spot; $M_D < M_S$)} \\
    &\text{iii) } && \lambda = -M_D: \quad 
    && \frac{y_2}{y_1} = 1
    .
    \quad \text{($Z$ blind spot)}
\end{alignat*}

To diagonalize the mass matrix, we can use Takagi diagonalization (see, e.g. Refs. \cite{Dreiner:2023yus, Dreiner:2008tw}).
\begin{equation}
    V^{*}\mathcal{M}_N V^{\dagger} = \text{diag}(m_1, m_2, m_3),
\end{equation}
with physical masses $m_i \in \mathbb{R}^{+}$ for $i = 1, 2, 3$. Our goal is to find the entries of the matrix $V$.

To find the entries of $V$, we note that, since in our case (i.e. no $CP$ violation) $\mathcal{M}_N$ is a real, symmetric matrix, we can find an orthogonal matrix $\mathcal{O}$ such that
\begin{equation} \label{eq:orthogdiagonalization}
    \mathcal{O}\mathcal{M}_N\mathcal{O}^{T} = \text{diag}(\epsilon_1 m_1, \epsilon_2 m_2, \epsilon_3 m_3)
    ,
\end{equation}
where physical masses $m_i \in \mathbb{R}^{+}$ and signs $\epsilon_i =\pm 1$.  If we can find such a matrix $\mathcal{O}$, then we can find the unitary matrix that diagonalizes $\mathcal{M}_N$ via
\begin{equation} \label{eq:unitarymix}
    V = \text{diag}(\sqrt{\epsilon_1}, \sqrt{\epsilon_2}, \sqrt{\epsilon_3})^{*}\mathcal{O}.
\end{equation}

Using Eq.~(\ref{eq:characteristiceq}), Eq.~(\ref{eq:orthogdiagonalization}), and the fact that $\mathcal{O}^{T}\mathcal{O} = I$, we can express $\mathcal{O}_{ij}$ in terms of $\epsilon_i$ and $m_i$:  
\begin{equation} \label{eq:Oij}
\begin{aligned}
    \mathcal{O}_{i3} &= \sqrt{\frac{y_1^2 v^2 + 2\epsilon_i m_i (M_S - \epsilon_i m_i)}{2M_D^2 + (y_1^2 + y_2^2)v^2 + 2 \epsilon_i m_i (2M_S - 3 \epsilon_i m_i)}},\\
    \mathcal{O}_{i2} &= \mathcal{O}_{i3} \frac{2M_D (M_S - \epsilon_i m_i) - y_1 y_2 v^2}{2 \epsilon_i m_i (M_S - \epsilon_i m_i) + y_1^2 v^2},\\
    \mathcal{O}_{i1} &= -\frac{v}{\sqrt{2}}\frac{y_1 \mathcal{O}_{i2} + y_2 \mathcal{O}_{i3}}{(M_S - \epsilon_i m_i)},
\end{aligned}
\end{equation}
where no summation over $i$ is implied in $\epsilon_i m_i$.

The couplings $g_{Z \chi^0_1\chi^0_1}$ and $g_{h \chi^0_1\chi^0_1}$, in terms of the DM mass $m_1$ and the input parameters, are given by \cite{Arcadi:2024ukq, Cohen:2011ec, Calibbi:2015nha, Cheung:2013dua}
\begin{eqnarray}
g_{Z \chi^0_1\chi^0_1} & = & \frac{m_Z v (y_1^2 - y_2^2)}{(y_1^2 + y_2^2) v^2 + 2 M_D^2 + 2 \epsilon_1 m_1 (2 M_S - 3 \epsilon_1 m_1)},\label{eq:Zcoupling}\\
g_{h \chi^0_1\chi^0_1} & = & \frac{-2 v \epsilon_1(2y_1y_2 M_D + (y_1^2 + y_2^2) \epsilon_1 m_1)}{(y_1^2 + y_2^2) v^2 + 2 M_D^2 + 2 \epsilon_1 m_1 (2 M_S - 3 \epsilon_1 m_1)}.\label{eq:Higgscoup}
\end{eqnarray}
Below, we report the couplings for a few cases of particular interest to this work.  

\subsection{Double blind spot}
\label{sec:appdoubleBS}

First, consider the double blind spot, $y_2 = -y_1$ when $M_D < M_S$ (see Eq.~(\ref{eq:DoubleBS})).
In this case, the physical masses are\footnote{In deriving these expressions we have also assumed that $0< y_1 < \frac{M_D}{v}\sqrt{2r_m(r_m - 1)}$.} 
\begin{equation}
    \begin{aligned}\label{eq:DoubleBS_MassEigenvalues}
        m_1 &= M_D,\\
        m_2 &= \frac{1}{2}\left(M_D - M_S + \sqrt{(M_S + M_D)^2 + 4y_1^2v^2}\right),\\
        m_3 &= \frac{1}{2}\left(M_S - M_D + \sqrt{(M_S + M_D)^2 + 4y_1^2v^2}\right),
    \end{aligned}
\end{equation}
with signs $\epsilon_1 = -\epsilon_2 = \epsilon_3 = 1$, such that the three eigenvalues $\lambda = \epsilon_i m_i$ (no summation over $i$) satisfy the characteristic Eq.~(\ref{eq:characteristiceq}) for $y_2 = - y_1$.

We also obtain the following values for the entries for $\mathcal{O}$ using Eq.~(\ref{eq:Oij})
\begin{equation}
    \begin{aligned}
        \mathcal{O}_{13} &= \mathcal{O}_{12} = \frac{1}{\sqrt{2}},\\
        \mathcal{O}_{23} &= -\mathcal{O}_{22} = \frac{y_1 v}{\sqrt{2(M_S + M_D)(M_S - m_3) + 4y_1^2v^2}},\\
        \mathcal{O}_{33} &= -\mathcal{O}_{32} = \frac{y_1 v}{\sqrt{2(M_S + M_D)(m_3 + M_D) + 4y_1^2v^2}},
    \end{aligned}
\end{equation}
and the entries $\mathcal{O}_{i1}$ can be determined from $\mathcal{O}_{i3}$ and $\mathcal{O}_{i2}$.

Using Eq.~(\ref{eq:SMcouplings}), we obtain the couplings of the dark sector to the SM Higgs and gauge bosons. Explicitly, the vector and axial $W$ couplings are
\begin{equation}\label{eq:Wcoups_doubleBS}
\begin{aligned}
&g_{W^{+} \chi^{-} \chi_1^0}^V=\frac{g}{2},
\quad g_{W^{+} \chi^{-} \chi_2^0}^V=\frac{i g}{\sqrt{2}} \mathcal{O}_{23},
\quad g_{W^{+} \chi^{-} \chi_3^0}^V=0,\\
&g_{W^{+} \chi^{-} \chi_1^0}^A=0,
\quad g_{W^{+} \chi^{-} \chi_2^0}^A=0,
\quad g_{W^{+} \chi^{-} \chi_3^0}^A=\frac{g}{\sqrt{2}} \mathcal{O}_{33}.
\end{aligned}
\end{equation}
The $Z$ couplings are:
\begin{equation}\label{eq:Zcoups_doubleBS}
\begin{aligned}
&g_{Z \chi_1^0 \chi_1^0} = g_{Z \chi_2^0 \chi_2^0} =
g_{Z \chi_3^0 \chi_3^0} = g_{Z \chi_2^0 \chi_3^0} = 0,\\
&g_{Z \chi_1^0 \chi_2^0} = \frac{i g}{\sqrt{2} c_w} \mathcal{O}_{23},
\quad g_{Z \chi_1^0 \chi_3^0}=\frac{g}{\sqrt{2} c_w} \mathcal{O}_{33}.
\end{aligned}
\end{equation}
The unsymmetrized couplings to the Higgs boson are:
\begin{equation}\label{eq:hcoups_doubleBS}
\begin{aligned}
&g_{h \chi_1^0 \chi_1^0}=g_{h \chi_1^0 \chi_2^0}=g_{h \chi_2^0 \chi_1^0}=g_{h \chi_1^0 \chi_3^0}=g_{h \chi_3^0 \chi_1^0}=0,\\
&g_{h \chi_2^0 \chi_2^0}=g_{h \chi_3^0 \chi_3^0}=\frac{2 y_1^2 v}{\left(2 m_3 + M_D - M_S\right)},\\
&g_{h \chi_2^0 \chi_3^0}=i \frac{2 y_1 \left(m_3 + M_D\right)}{\left(2 m_3 + M_D - M_S\right)},\\
&g_{h \chi_3^0 \chi_2^0}=-i \frac{2 y_1 \left(m_3 - M_S\right)}{\left(2 m_3 + M_D - M_S\right)}.
\end{aligned}
\end{equation}

\subsection{Small Yukawa couplings}
\label{sec:appSmally}

Now we turn to the small Yukawa coupling limit.
In this regime, it is particularly convenient to express the mass mixing of the neutral states, as given in Eqs.~(\ref{eq:Lagrangian}) and (\ref{eq:mixingmatrix}), in terms of the rotated gauge eigenstate basis
$\widetilde{\psi}^0 \equiv (S, \frac{\overline{N} + N}{\sqrt{2}}, \frac{\overline{N} - N}{\sqrt{2}})$:
\begin{equation} \label{eq:rotatedbasis}
    \mathcal{L}_{mass} \supset
    -\frac{1}{2}(\widetilde{\psi}^0)^T
    \begin{pmatrix}
        M_S & \frac{y_+ v}{2} & -\frac{y_- v}{2}\\
        \frac{y_+ v}{2} & M_D & 0 \\
        -\frac{y_- v}{2} & 0 & -M_D
    \end{pmatrix}
    \widetilde{\psi}^0
    ,
\end{equation}
where the small Yukawa coupling combinations $y_\pm = y_1 \pm y_2$ appear off-diagonally, and the mass parameters $M_S$ and $M_D$, which are much larger than $\left|y_\pm v\right|$, appear on the diagonal.

We expand the mass eigenvalues and the orthogonal matrix that diagonalizes the mass matrix above (via Eq.~(\ref{eq:orthogdiagonalization})) in powers of $v$ accompanying the small Yukawa couplings, and solve iteratively up to order $\mathcal{O}(v^2)$.
The physical masses of the neutral states are
\begin{equation}
\label{eq:SmallY_MassEigenvalues}
    \begin{aligned}
        m_a &= M_S - \frac{v^2 y_+^2}{4 \left(M_D - M_S\right)} + \frac{v^2 y_-^2}{4 \left(M_D + M_S\right)},\\
        m_b &= M_D + \frac{v^2 y_+^2}{4\left(M_D - M_S\right)},\\
        m_c &= M_D + \frac{v^2 y_-^2}{4\left(M_D + M_S\right)},
    \end{aligned}
\end{equation}
with signs $\epsilon_a = \epsilon_b = -\epsilon_c = 1$,
and the unitary matrix that diagonalizes the mass matrix is determined from Eq.~(\ref{eq:unitarymix}).
Here, we have used $a,b,c$ to label the three eigenvalues because they are not necessarily in order-- the ordering depends on, e.g., the relative size of $M_{S}$ and $M_{D}$.
When $M_S < M_D$, then we have $a = 1$.  When $M_D < M_S$, we have $b=1$.
 
 In the small $y$ limit, the vector and axial $W^{+}-\chi^{-}-\chi^{0}_a$ couplings are given by:
\begin{equation} \label{eq:Wcoups_smallY}
\begin{aligned}
g_{W^{+} \chi^{-} \chi_a^0}^V=-\frac{g}{4} \frac{vy_+}{\left(M_D-M_S\right)},&\quad  
g_{W^{+} \chi^{-} \chi_a^0}^A=-\frac{g}{4} \frac{vy_-}{\left(M_D+M_S\right)},\\
g_{W^{+} \chi^{-} \chi_b^0}^V=\frac{g}{2}\left[1-\frac{v^2y_+^2}{8\left(M_D-M_S\right)^2}\right],&\quad
g_{W^{+} \chi^{-} \chi_b^0}^A=-\frac{g}{2} \frac{v^2y_+y_-}{8 M_D\left(M_D-M_S\right)},\\
g_{W^{+} \chi^{-} \chi_c^0}^V=i \frac{g}{2}\left[1-\frac{v^2y_-^2}{8\left(M_D+M_S\right)^2}\right],&\quad
g_{W^{+} \chi^{-} \chi_c^0}^A=-i \frac{g}{2} \frac{v^2y_+y_-}{8 M_D\left(M_D+M_S\right)}.
\end{aligned}
\end{equation}
Note that the magnitude of both $g_{W^{+} \chi^{-} \chi_b^0}^V$ and $g_{W^{+} \chi^{-} \chi_c^0}^V$ to leading order are $g/2$ even in the small Yukawa couplings as they connect the charged and neutral part of the $SU(2)_{L}$ doublet.

The couplings of the neutral states to the $Z$ boson are:
\begin{equation}
\label{eq:Zcoups_smallY}
\begin{aligned}
& g_{Z \chi_a^0 \chi_a^0}=\frac{g}{4 c_w} \frac{v^2y_+y_-}{\left(M_D^2-M_S^2\right)}, \\
& g_{Z \chi_b^0 \chi_b^0}=-\frac{g}{8 c_w} \frac{v^2y_+y_-}{M_D\left(M_D-M_S\right)}, \\
& g_{Z \chi_c^0 \chi_c^0}=-\frac{g}{8 c_w} \frac{v^2y_+y_-}{M_D\left(M_D+M_S\right)}, \\
& g_{Z \chi_a^0 \chi_b^0}=-\frac{g}{4 c_w} \frac{vy_-}{\left(M_D+M_S\right)}=g_{Z \chi_b^0 \chi_a^0}^*, \\
& g_{Z \chi_b^0 \chi_c^0}=i \frac{g}{2 c_w}
\left[1-
\frac{v^2 y_+^2}{8\left(M_D - M_S\right)^2} -
\frac{v^2 y_-^2}{8\left(M_D + M_S\right)^2}
\right]=g_{Z \chi_c^0 \chi_b^0}^*, \\
& g_{Z \chi_a^0 \chi_c^0}=-i \frac{g}{4 c_w} \frac{vy_+}{\left(M_D-M_S\right)}=g_{Z \chi_c^0 \chi_a^0}^*.
\end{aligned}
\end{equation}
Note that $g_{Z \chi_b^0 \chi_c^0}$ to leading order has magnitude $g/2c_w$ in the small Yukawa coupling limit, as this coupling connects $SU(2)_L$ doublet states.

The unsymmetrized couplings to the Higgs boson are
\begin{equation} \label{eq:hcoups_smallY}
\begin{aligned}
 g_{h \chi_a^0 \chi_a^0}&=-\frac{v y_+^2}{2\left(M_D - M_S\right)} + \frac{v y_-^2}{2\left(M_D + M_S\right)}, \\
 g_{h \chi_b^0 \chi_b^0}&=\frac{vy_+^2}{2\left(M_D-M_S\right)}, \\
 g_{h \chi_c^0 \chi_c^0}&=\frac{vy_-^2}{2\left(M_D+M_S\right)}, \\
 g_{h \chi_a^0 \chi_b^0}&=-\frac{v^2y_+}{4\left(M_D-M_S\right)}\left[\frac{y_+^2}{\left(M_D - M_S\right)} - \frac{y_-^2}{\left(M_D + M_S\right)}\right], \\
 g_{h \chi_b^0 \chi_c^0}&=i \frac{vy_+y_-}{2\left(M_D+M_S\right)}, \\
 g_{h \chi_a^0 \chi_c^0}&=-i \frac{v^2y_-}{4\left(M_D+M_S\right)}\left[\frac{y_+^2}{\left(M_D - M_S\right)} - \frac{y_-^2}{\left(M_D + M_S\right)}\right], \\
 g_{h \chi_b^0 \chi_a^0}&= y_+
 \left[
 1 -
 \frac{v^2 y_+^2}{4 \left(M_D - M_S\right)^2} +
 \frac{v^2 y_-^2}{8 \left(M_D + M_S\right)^2} \frac{M_S \left(3 M_D + M_S\right)}{M_D \left(M_D - M_S\right)}
 \right],\\
 g_{h \chi_c^0 \chi_b^0}&=-i \frac{vy_+y_-}{2\left(M_D-M_S\right)}, \\
 g_{h \chi_c^0 \chi_a^0}&= -i y_-
 \left[
 1 -
 \frac{v^2 y_-^2}{4 \left(M_D + M_S\right)^2} -
 \frac{v^2 y_+^2}{8 \left(M_D - M_S\right)^2} \frac{M_S \left(3 M_D - M_S\right)}{M_D \left(M_D + M_S\right)}
 \right].
\end{aligned}
\end{equation}
Note that the couplings
$g_{h \chi_b^0 \chi_a^0}$ and $g_{h \chi_c^0 \chi_a^0}$ to leading order are $y_+$ and $y_-$, respectively, in the small Yukawa coupling limit.
These couplings connect the $SU(2)_L$ doublet states $\frac{\overline{N} + N}{\sqrt{2}}$ and $\frac{\overline{N} - N}{\sqrt{2}}$ in the rotated gauge eigenstate basis to the $SU(2)_L$ singlet state $S$.

\bibliographystyle{apsrev4-1}
\bibliography{ref}

\end{document}